\documentclass[aps,pra,superscriptaddress,twocolumn]{revtex4-2}

\usepackage{graphicx}
\usepackage{amsmath}
\usepackage{amssymb}
\usepackage{hyperref}
\usepackage[utf8]{inputenc}
\usepackage{mathtools}
\usepackage[english]{babel}
\usepackage{bbm}
\hypersetup{colorlinks=true, linkcolor=blue, citecolor=blue, urlcolor=blue}
\usepackage{braket}
\usepackage[normalem]{ulem}

\usepackage[dvipsnames]{xcolor}
\newcommand{\new}[1]{\textcolor{Black}{#1}}

\newcommand{\ie}{i.\,e.,\ }

\newcommand{\eg}{e.\,g.,\ }

\newcommand{\re}{\mathrm{Re}}
\newcommand{\im}{\mathrm{Im}}

\newcommand{\rr}{\mathbf{r}}

\newcommand{\fref}[1]{\text{Fig.}~\ref{#1}}

\newcommand{\eref}[1]{\text{Eq.}~\eqref{#1}}

\begin{document}
\title{Quantum computing with subwavelength atomic arrays}
\author{Freya Shah}
\affiliation{Department of Physics, Harvard University, Cambridge, Massachusetts 02138, USA}
\affiliation{School of Arts and Sciences, Ahmedabad University, Ahmedabad 380015, Gujarat, India}
\author{Taylor L. Patti}
\affiliation{Department of Physics, Harvard University, Cambridge, Massachusetts 02138, USA}
\affiliation{NVIDIA, Santa Clara, California 95051, USA}
\author{Oriol Rubies-Bigorda}
\affiliation{Physics Department, Massachusetts Institute of Technology, Cambridge, Massachusetts 02139, USA}
\affiliation{Department of Physics, Harvard University, Cambridge, Massachusetts 02138, USA}
\author{Susanne F. Yelin}
\affiliation{Department of Physics, Harvard University, Cambridge, Massachusetts 02138, USA}

\begin{abstract}

Photon-mediated interactions in subwavelength atomic arrays have numerous applications in quantum science. In this manuscript, we explore the potential of three-level quantum emitters, or ``impurities" embedded in a two-dimensional atomic array to serve as a platform for quantum computation. By exploiting the altered behavior of impurities as a result of the induced dipole-dipole interactions mediated by subwavelength array, \new{we design and simulate a set of universal quantum gates} consisting of the $\sqrt{\text{iSWAP}}$ and single-qubit rotations. \new{We demonstrate that these gates have very high fidelities due to the long atomic dipole-dipole coherence times}, as long as the atoms remain within a proximal range. 
Finally, \new{we design and simulate quantum circuits leading to the generation of the maximally entangled two-qubit Bell states}, as well as the entangled three-qubit GHZ state. These findings establish subwavelength emitter arrays as an
alternative platform for quantum computation and quantum simulation.

\end{abstract}

\maketitle


\section{Introduction}

\begin{figure}
\centering
\includegraphics[width=0.5\textwidth]{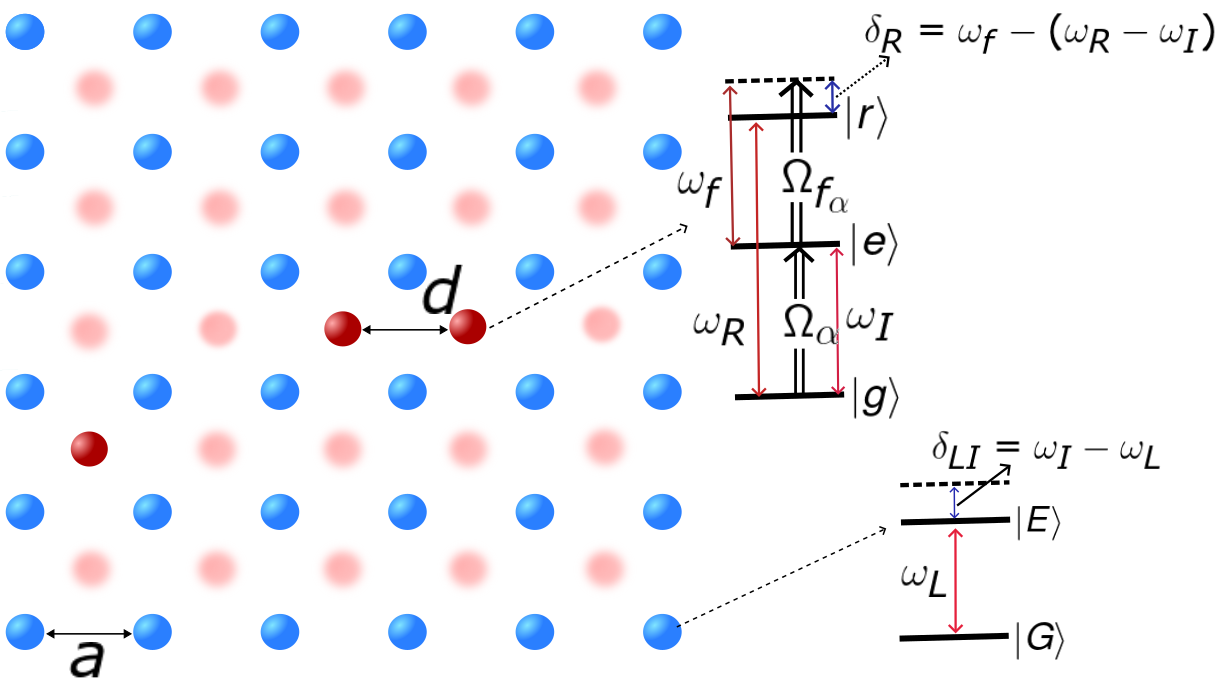}
\caption{Square atomic array with subwavelength spacing $a$ and with embedded impurities (red/dark grey) at the center of the square plaquette. The impurity emitters are three-level systems. The $|g \rangle \leftrightarrow |e\rangle$ transition exhibits light-induced dipole-dipole interactions with the lattice atoms and the remaining impurities. The third level $|r \rangle$ is a metastable state with a small decay rate. The lattice atoms (blue/light grey) are two-level emitters with transition frequency detuned by $\delta_{LI}$ with respect to that of the impurity atoms. }
\label{fig:model}
\end{figure}

Recent experimental advances in atomic array configurations with subwavelength spacing and long-range dipole-dipole interactions \cite{rui2020subradiant} have opened up exciting possibilities for exploring novel physical phenomena, such as superradiance \cite{dicke1954coherence,masson2022universality,sierra2022dicke, super_Robi,rubies2022superradiance,rubies2023characterizing} and subradiance \cite{bienaime2012controlled,asenjo2017exponential,rubies2022generating,Cosimo_subradiance,subradiance_rings_mariona}, topological quantum optics \cite{Janos_topo,topo_Bettles}, efficient light-matter interactions \cite{Shahmoon2017,reitz2022cooperative,mirror_adams,Ruoste_wavefront,walther_rydberg} and enhanced photon storage capabilities \cite{facchinetti2016storing,asenjo2017exponential,ballantine2021quantum,Rubiesbigorda2022}.
Additionally, atomic arrays can be employed to modify the radiative environment and properties of impurity emitters, illustrated by the red atoms in Fig.~\ref{fig:model}. In that case, the array acts as a Markovian bath for the embedded impurities, which acquire a suppressed decay rate and exhibit long-range interactions mediated by the delocalized spin waves of the lattice \cite{Patti2021,Masson2020,Malz_impurities,Samuel_impurities}.

In this manuscript, we explore the potential of subwavelength atomic arrays with embedded impurity emitters as candidates for quantum computation. Though relatively novel computational paradigm, with applications ranging from quantum simulation \cite{Georgescu2014,Lloyd1996,Feynman2018} to cryptography \cite{Gisin2002,Pirandola2020} and optimization \cite{Farhi2014,Peruzzo2014,Li2020,patti2022variational}. 
\new{However, the realization of universal quantum computation, that is, the ability to perform any combination of quantum operations efficiently \cite{Campbell2017} using highly scalable systems with
exceptionally low errors, has been an elusive and long-standing goal \cite{Divincenzo2000,Ladd2010}. To address this difficulty, a multitude of quantum hardware platforms have been proposed. Of these paradigms, superconducting qubits have consistently ranked among the most popular and have attained numerous noteworthy milestones \cite{Chow2012,Devoret2013,Macklin2015,Neill2018}, however error mitigation on these platforms remains an engineering challenge \cite{Kjaergaard2020,Siddiqi2021,Bravyi2022}. Trapped ion systems are remarkable for their implementation in free space and their long coherence times \cite{Cirac2000,Bruzewicz2019}, however their utilization of motional quantum states presents unique challenges for their design and scalability \cite{Brown2021}. Neutral atoms have garnered increasing interest in the quantum computing community due to their relatively scalable configuration, which sports high uniformity and reproducibility \cite{Jaksch2000,Henriet2020}, although quantum hardware efforts have so far been largely limited to cold Rydberg atoms in optical arrays \cite{Bernien2017,Adams2019,Browaeys2020,Bluvstein2021}.} While each platform offers unique advantages, they also present their own limitations and challenges (\eg decoherence, sensitivity to external noise, and scalability with increasing qubit number), which have slown the path towards quantum computing's full potential \cite{Alfieri2022,Cheng2023}.

\new{Here, we theoretically and numerically demonstrate that impurities embedded in closely-spaced atomic arrays can serve as qubits suitable for quantum computation} and simulation protocols.
For that, we show that the lattice-mediated interactions implement an iSWAP gate between the impurities.
Individual addressing of each impurity further allows to suppress these interactions and thereby enables precise control over the iSWAP gate operation, as well as the implementation of the single-qubit X and Z rotations. Together, these three operations form a universal set of gates for quantum computation and simulation.
\new{Due to the low error rates of these operations}, they can be applied sequentially to generate quantum circuits with large numbers of gates and qubits or impurities.

\new{While efforts to develop the useful quantum computing platforms are multi-faceted and their challenges ever-evolving, our work demonstrates the remarkable attributes of lattice-embedded atomic emitters, such as their coherent controllability. The ability to interact with 3D free space light and generate optical lattices makes it a useful tool for quantum sensing and simulation. Moreover, due to the high controllability, the utilization of impurities can be instrumental in applications like direct manipulation and retrieval of quantum data, and generation of cluster states for quantum communication and quantum internet. It can also be used for entanglement generation, making single-photon sources, and gaining control and access to dark lattice modes~\cite{Rubiesbigorda2022}. Indeed, all of these advantageous prospects are complemented by the fascinating effects that these quantum emitters demonstrate.}

The rest of the work is organized as follows. In Section \ref{section:model}, we introduce the system and derive the effective equations of motion of the embedded impurities. In Section \ref{section:gates}, \new{we derive the feasibility of the aforementioned universal gate set. We further illustrate the feasibility of quantum circuits in this system by preparing a two-qubit Bell state and the three-qubit Greenberger–Horne–Zeilinger (GHZ) state} in Section \ref{section:operations}. These states serve as the universal computational primitive \cite{gottesman1999quantum} and thus highlight the robustness and efficacy of our platform for implementing various quantum algorithms with high precision.

\section{Model}
\label{section:model}

We consider a two-dimensional array of $N_{L}$ two-level atoms with $N_I$ embedded three-level impurity atoms, as depicted by the blue and red emitters in Fig.~\ref{fig:model}, respectively. The two levels of the array emitters are labeled as $|G\rangle$ and $|E\rangle$, and \new{the angular frequency of their transition is $\omega_L = 2 \pi c/\lambda_{L}$. As all frequencies in this manuscript are angular frequencies, we will henceforth unambiguously refer to them as ``frequencies''.} We additionally consider a closely-spaced lattice, such that the lattice spacing $a < \lambda_{L}$. The three levels of the impurity atoms are the ground state $\ket{g}$, the excited state $\ket{e}$ and the high-energy metastable state (HMS) $\ket{r}$, as shown in ~\fref{fig:model}. The frequency of the $\ket{g} \leftrightarrow \ket{e}$ and  $\ket{g} \leftrightarrow \ket{r}$ are $\omega_I$ and $\omega_R$, respectively. 

\new{Both the lattice atoms and the impurities interact with the vacuum electromagnetic field. Applying the Born-Markov approximation, we trace out the electromagnetic vacuum radiation field and obtain the master equation describing the density operator of the emitters. In the quantum jump formalism, the atomic dynamics are equally described by an effective non-Hermitian Hamiltonian and the stochastic action of quantum jumps, responsible for photon emission. In what follows, we assume that the action of the jumps can be neglected and the dynamics of the system is fully characterized by the non-Hermitian Hamiltonian $H=H_{L}+H_{I}+H_{LI}+H_{E}$. As this effective non-Hermitian Hamiltonian provides a full description of our quantities of interest, namely atomic dynamics, we present it here for simplicity, however we also offer its full derivation in Appendix \ref{appdx: master eqn}. Then, the lattice Hamiltonian  $\hat{H}_L$ can be written as}
\begin{equation}
     \hat{H}_{\text{L}}= \sum_{i}^{N_{L}}(w_{L}-\frac{i}{2}\gamma _{L})\hat{\sigma} _{i}^{\dagger }\sigma _{i}+\sum_{i,j\neq i}^{N_{L}}(J_{ij}-\frac{i}{2}\Gamma_{ij} )\sigma _{i}^{\dagger }\sigma _{j},
     \label{eq: Hamiltonian_lattice}
\end{equation}
where $\gamma_{L}$ is the spontaneous decay rate of the lattice atoms and $\hat{\sigma}_i=\ket{G_i}\bra{E_i}$ is the lowering operator for lattice atom $i$. The eliminated vacuum field induces coherent and dissipative interactions between the emitters, which arise from virtual emission and reabsorption of photons. These couplings $J_{ij} \equiv J_{ij}(\mathbf{r}_{i},\mathbf{r}_{j})$ and $\Gamma_{ij} \equiv \Gamma_{ij}(\mathbf{r}_{i},\mathbf{r}_{j})$  depend on the distance $r_{ij}=|\mathbf{r}_{i}-\mathbf{r}_{j}|$ between atoms $i$ and $j$, and are given by
\begin{subequations}
    \begin{align}
    J_{ij}(\rr_i,\rr_j) &= -\frac{3\pi \gamma_L}{w_{L}} {\mathbf{d}_i^\dagger} \cdot \re [\textbf{G}(\textbf{r}_{ij}, w_{L})] \cdot \mathbf{d}_j, \\
    \Gamma_{ij}(\rr_i,\rr_j) &= \frac{6\pi \gamma_L}{w_{L}} {\mathbf{d}_i^\dagger} \cdot \im [\textbf{G}(\textbf{r}_{ij},w_{L})] \cdot \mathbf{d}_j.
    \end{align}
\end{subequations}
Here, $\textbf{G}(\textbf{r},w) $ is the dyadic Green's tensor \cite{gruner1996green,dung2002resonant}, and $\mathbf{d}_{i}$ is the atomic dipole moment of atom $i$. 

Similarly, the impurity Hamiltonian $\hat{H}_I$ reads
\begin{align}
\label{eq: Hamiltonian_impurity}
    H_{\text{I}} &= \sum_{\alpha}^{N_{I}}\Bigl[(w_{I}-\frac{i}{2}\gamma _{I})s_{\alpha}^{\dagger}s_{\alpha} + (w_{R}-\frac{i}{2}\gamma _{R})r_{\alpha}^{\dagger}r_{\alpha}\Bigr] \nonumber \\
    &+\sum_{\alpha,\beta\neq \alpha}^{N_{I}}\frac{\gamma_{I}}{\gamma_{L}} (J_{\alpha\beta}-\frac{i}{2}\Gamma_{\alpha\beta})s_{\alpha}^{\dagger}s_{\beta},
\end{align}
where $\gamma_I$ and $\gamma_R$ are the spontaneous decay rates of excited and HMS states of the impurities, and $s_\alpha=\ket{g_\alpha}\bra{e_\alpha}$ and $r_\alpha=\ket{e_\alpha}\bra{r_\alpha}$ are the lowering operators of the $\ket{g} \leftrightarrow \ket{e}$ and $\ket{e} \leftrightarrow \ket{r}$ transitions, respectively. We consider only the $\ket{g} \leftrightarrow \ket{e}$ transition to have a transition wavelength $\lambda_I = 2 \pi c / \omega_I$ of the order of the impurity distance. Thus, only this transition exhibits significant light-induced couplings $J_{\alpha\beta}$ and $\Gamma_{\alpha\beta}$. Thus, the HMS states of different impurities do not interact with one another and can therefore be used to store the quantum state of the impurities for long times provided that $\gamma_R \ll \gamma_I$.

\new{Provided that $\omega_L \approx \omega_I$, impurity and lattice atoms experience the same near-resonant atom-atom coupling that lattice atoms do with each other, as described by Eq.\ \ref{eq: Hamiltonian_lattice} and Heisenberg-Langevin equation derivation in Appendix \ref{appdx: master eqn}. That is, the lattice and impurity emitters also undergo light-induced interactions given by}
\new{\begin{align}
    \begin{split}
     H_{\text{LI}}&=\sqrt{\frac{\gamma_{I}}{\gamma_{L}}} \sum_{i}^{N_{L}}\sum_{\alpha}^{N_{I}}\Bigl [(J_{i\alpha}-\frac{i}{2}\Gamma_{i\alpha})\sigma_{i}^{\dagger}s_{\alpha}+(J_{\alpha i}-\frac{i}{2}\Gamma_{\alpha i})s_{\alpha}^{\dagger}\sigma_{i}\Bigr ],
     \label{eq: Hamiltonian_lattice_impurity}
    \end{split} 
\end{align}}

\noindent \new{where the sole distinction between the atomic couplings in Eq.\ \ref{eq: Hamiltonian_lattice} and Eq.\ \ref{eq: Hamiltonian_lattice_impurity} is the relative coupling strength $\sqrt{\gamma_\text{I}/\gamma_\text{L}}$ as dictated by the relative linewidth of the impurity and lattice atoms, respectively.}

Finally, \new{classical driving fields can be applied} on the $\ket{g} \leftrightarrow \ket{e}$ and $\ket{e} \leftrightarrow \ket{r}$ transitions of the impurity atoms with frequencies $\omega_I$ and $\omega_f$, respectively,
\begin{align}
\begin{split}
    H_{\text{E}}&= -\sum_{\alpha}^{N_{I}} \Bigl[ (\Omega_\alpha^* s_{\alpha}^{\dagger}+ \Omega_\alpha s_{\alpha})(e^{-i\omega_{I}t} +e^{i\omega_{I}t}) \\&+ (\Omega_{f \alpha}^* r_{\alpha}^{\dagger}+\Omega_{f \alpha} r_{\alpha})(e^{-i\omega_{f} t} +e^{i\omega_{f} t})\Bigr],
   \end{split} 
\end{align}
where the \new{position-dependent} strengths \new{$\Omega_{\alpha}= -\mathbf{d}_{\alpha} \boldsymbol{\varepsilon}_{\alpha}$ and $\Omega_{f \alpha}= -\mathbf{d}_{\alpha f}\boldsymbol{\varepsilon}_{f \alpha}$} are the product of the corresponding dipole moments and \new{the} electric field \new{at the atomic positions}. \new{Here, the position-dependent phases are absorbed into the strengths $\Omega_{\alpha}$ and $\Omega_{f \alpha}$} ~\cite{Scully1997}.

For simplicity, we further consider that the system contains at most two excitations. Applying the Schrödinger equation, we can obtain the equations of motion of the amplitudes in each of the states of the lattice and impurity atoms, derived in Appendix~\ref{appendix:EOMS}. If $\gamma_{L} \gg \gamma_{I}$, the lattice atoms act as a Markovian bath for the impurities and can therefore be adiabatically eliminated, as shown in Ref.~\cite{Patti2021} and derived in Appendix~\ref{appendix:EOMS}. We then obtain a reduced set of equations for the $N_I$ impurity emitters only. Their wavefunction is then simply given by 
\begin{widetext}
\begin{align}
\label{eqn:6}
    \left|\Psi(t)\right \rangle &= a(t)\left|g\right \rangle +  \sum_{\alpha}^{N_{I}}c_{\alpha}(t)e^{-iw_{I}t}\left|e_{\alpha}\right \rangle + \sum_{\alpha}^{N_{I}}f_{\alpha}(t)e^{-i(w_{I}+w_{f})t}\left|r_{\alpha}\right \rangle +
    \sum_{\alpha,\beta\neq \alpha}^{N_{I}}C_{\alpha \beta}(t)e^{-2iw_{I}t}\left|e_{\alpha}, e_{\beta}\right \rangle \nonumber \\ &+ \sum_{\alpha,\beta\neq \alpha}^{N_{I}}y_{\alpha \beta}(t)e^{-i(2w_{I}+w_{f})t}\left|e_{\alpha}, r_{\beta}\right \rangle +
    \sum_{\alpha,\beta \neq \alpha}^{N_{I}} F_{\alpha \beta}(t)e^{-i(2w_{I}+2w_{f})t}\left|r_{\alpha}, r_{\beta}\right \rangle,
\end{align}
\end{widetext}

\noindent where $\ket{g}$ denotes the state where all impurities are in ground state, $\ket{e_\alpha}$ the state where only impurity $\alpha$ is in the excited state, $\ket{e_\alpha, e_\beta}$ the state where only impurities $\alpha$ and $\beta$ are in the excited state, and similarly for $\ket{r_\alpha}$, $\ket{r_\alpha, r_\beta}$ and $\ket{e_\alpha, r_\beta}$. The corresponding equations of motion applying the rotating wave approximation are 

 \begin{widetext}
\begin{subequations}
\label{eqn:reduced_EOMS}
\begin{align}
    \dot{a}=\;&i\Omega_{\alpha } c_{\alpha}(t),\\
    \dot{c_{\alpha}}(t)=\;& i\Omega_{\alpha}^{*}a(t) +i[\frac{i}{2}\gamma_{I}-\Sigma_{\alpha}]c_{\alpha}(t)-i \sum_{\beta \neq \alpha}^{N_{I}} [\Phi_{\alpha \beta}+\phi_{\alpha \beta}] c_{\beta}(t)+i\Omega_{\beta}C_{\alpha\beta}(t)+i\Omega _{f_{\alpha}}f_{\alpha}(t),\\
    \dot{C}_{\alpha\beta}(t)=\;&i[i\gamma_{I}-\Sigma_{\alpha}-\Sigma_{\beta}]C_{\alpha\beta}(t)+i\Omega_{\beta}^{*}c_{\alpha}(t)+i\Omega_{\alpha}^{*}c_{\beta}(t)+i\Omega_{f \alpha}y_{\beta\alpha}(t)+i\Omega_{f \beta}y_{\alpha\beta}(t) \nonumber \\
    & -i \sum_{\epsilon \neq \alpha,\beta}^{N_{I}} \left( [\Phi_{ \beta \epsilon}+\phi_{\beta \epsilon}] C_{\alpha\epsilon} + [\Phi_{ \alpha \epsilon}+\phi_{\alpha \epsilon}] C_{\beta\epsilon}  \right), \\
    \dot{f}_{\alpha}(t)=\;& i(\delta_{R}+\frac{i}{2}\gamma _{R})f_{\alpha}(t) +i\Omega_{\beta}F_{\alpha\beta}(t) +\Omega_{f \alpha}^{*}c_{\alpha}(t),\\
    \dot{y}_{\alpha\beta}(t)=\;&i[\delta_{R}+\frac{i}{2}\gamma_{I}+\frac{i}{2}\gamma_{R}-\Sigma _{\alpha}]y_{\alpha\beta}(t)+i\Omega_{\beta}^{*}C_{\alpha\beta}(t) + i \Omega_{f_\alpha} F_{\alpha \beta} +i \Omega^*_\alpha f_\beta -i \sum_{\epsilon \neq \alpha,\beta}^{N_{I}} [\Phi_{ \alpha \epsilon}+\phi_{\alpha \epsilon}]  y_{\epsilon \beta}, \\
    \dot{F}_{\alpha \beta}(t)=\;& 2i(\delta_{R}+\frac{i}{2}\gamma_{R})F_{\alpha\beta}(t)+i\Omega_{f \alpha}^{*}y_{\alpha\beta}(t)+i\Omega_{f \beta}^{*}y_{\beta\alpha}(t),
    \end{align}
\end{subequations} 
\end{widetext}
Here, $\Omega_{\alpha}$ and $\Omega_{f_{\alpha}}$ denote the driving fields that excite impurity $\alpha$ from the ground state to the excited state, and from the excited state to the HMS state, respectively. As illustrated in Fig.~\ref{fig:model}, we introduce $\delta_{R}=\omega_{f}- (\omega_{R}- \omega_{I})$. The term 
\begin{equation}
    \Phi_{\alpha \beta} = \frac{\gamma_{I}}{\gamma_{L}}(J_{\alpha\beta}-\frac{i}{2}\Gamma_{\alpha\beta})
\end{equation}
describes the coherent and dissipative exchange of excitations between impurities $\alpha$ and $\beta$ mediated by the electromagnetic field, and its value is therefore $\sim \gamma_I$. Additionally, the lattice also mediates interactions between impurities. Defining the light-induced interaction between impurity $\alpha$ and the lattice atoms as $C_{\alpha}$, and the light-induced coupling between the lattice atoms as $L$, the lattice-mediated interactions between both impurities, $\phi_{\alpha,\beta}$, reads
\begin{subequations}
\label{eqn:LandC_define}
    \begin{align}
    \phi_{\alpha \beta} &= C_{\beta}^\dagger L^{-1} C_{\alpha}, \\
    C_{\alpha}^\dagger&=\sqrt{\frac{\gamma_{I}}{\gamma_{L}}} \left( J_{\alpha1}-\frac{i}{2}\Gamma_{\alpha1},\cdots, J_{\alpha N}-\frac{i}{2}\Gamma_{\alpha N} \right), \\
    L &=
    \begin{bmatrix}
        \delta_{LI}+\frac{i}{2}\gamma_{L} & \cdots   &-J_{1N}+\frac{i}{2}\Gamma_{1N}\\ 
        \vdots &\ddots &\vdots  \\ 
        -J_{N1}+\frac{i}{2}\Gamma_{N1}& \cdots  & \delta_{LI}+\frac{i}{2}\gamma_{L}
    \end{bmatrix} .
    \end{align}
\end{subequations}
where $\delta_{LI} = \omega_I - \omega_L$ denotes the detuning between the lattice and impurity atoms. That is, $\phi_{\alpha,\beta}$ describes an excitation transferred between both impurities through the normal modes of the lattice.

Finally, $\Sigma_{\alpha} = \phi_{\alpha \alpha}$ denotes the self-interaction of impurity $\alpha$ mediated by the lattice. Its real part, $\text{Re}[\Sigma_{\alpha}]$, modifies the transition frequency of the impurities. Since this term is equal for all impurities provided that the lattice is large enough, it can be simply canceled out by redefining the frequency of the impurities. The imaginary part, however,  alters the effective decay rate of the impurity emitters
\begin{equation}    \Gamma_{\text{eff}}=\gamma_{I}-2\text{Im}[\Sigma_{\alpha}].
\end{equation}
That is, we can reduce the impurity decay rate and thus improve the coherence time when $\text{Im}[\Sigma_{\alpha}]>0$. If all emitters are circularly polarized ($\mathbf{d}_{L}=\mathbf{d}_{I}=(1,i,0)/\sqrt{2}$), there exists an optimal detuning $\delta_{LI}$ between the impurity and lattice atoms for which $\Gamma_{\text{eff}} \ll \gamma_I$ and the impurities become long-lived \cite{Patti2021,Samuel_impurities}. In that case, the frequency of the impurities lies outside the energy band of the normal modes of the lattice and the impurity-impurity interactions are off-resonantly coupled by their guided, or non-decaying, modes. Crucially, this suppressed decay rate allows for longer times ($ \sim \Gamma_\mathrm{eff}^{-1}$) to perform quantum operations or gates, thereby increasing the maximum circuit-depth. As a result, we consider this configuration for the rest of this work. \new{Note that it is precisely the existence of this regime characterized by suppressed photon emission that allows to neglect the action of the quantum jumps. For times smaller than the inverse effective decay rate ($t \ll \Gamma_\mathrm{eff}^{-1}$), \ie for the relevant timescales where high-fidelity operations can be performed, the dynamics can be simply computed using the non-Hermitian Hamiltonian and a wavefunction, instead of the full master equation and the atomic density matrix. }

\section{Universal set of gates}
\label{section:gates}

A universal set of gates is a group of operations that can implement any unitary transformation on a given number of qubits. For an ensemble of impurity atoms or qubits coupled to the atomic array, \new{we devise a universal gate set comprised of arbitrary single-qubit operations}, using X and Z rotations, and the two-qubit gate $\sqrt{\text{iSWAP}}$ ~\cite{schuch2003natural}.

\subsection{X rotation}

\begin{figure}
\centering
\includegraphics[width = \columnwidth]{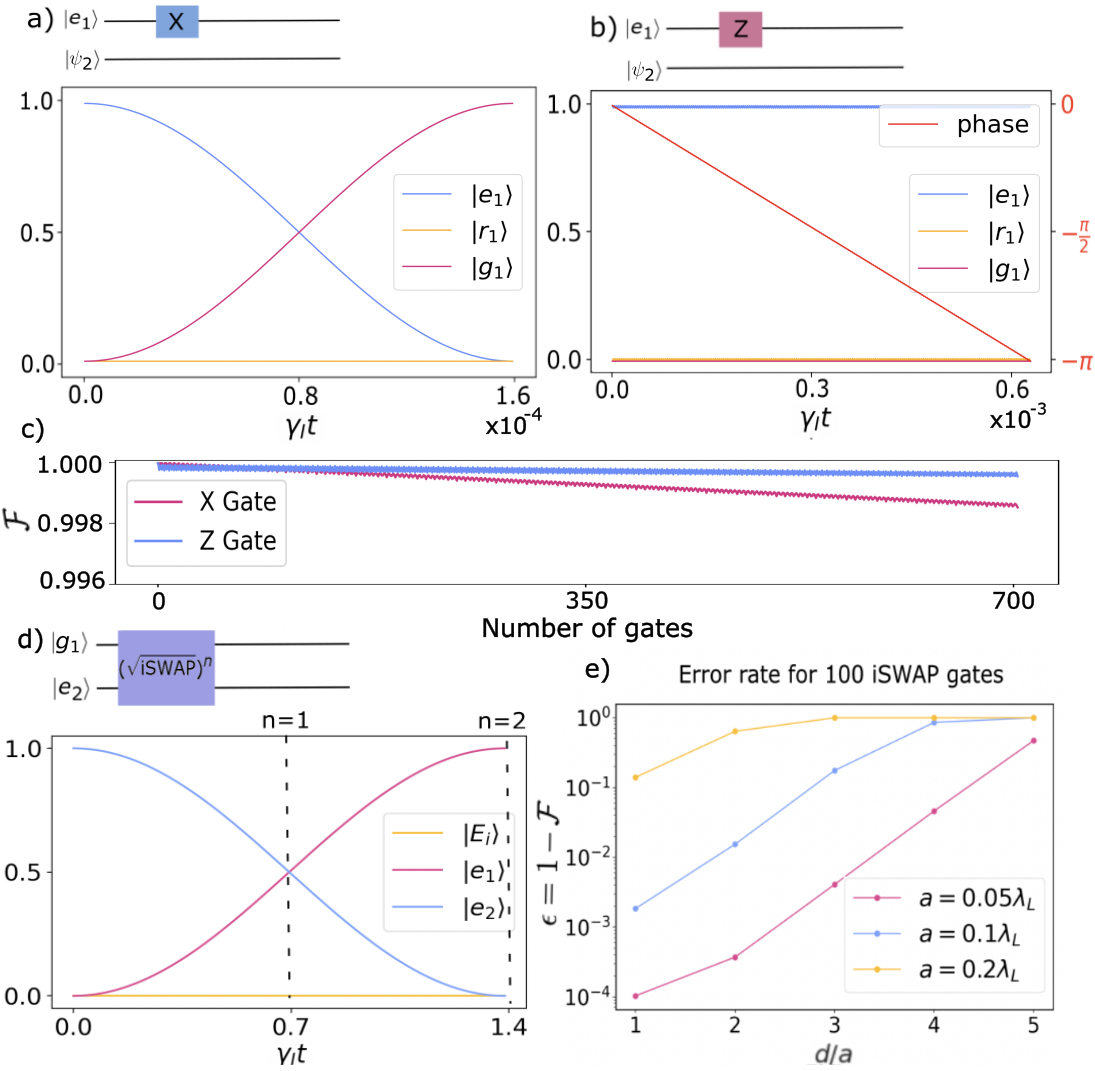}
\caption{ In the subfigures (a),(b),(c), and (d) we have considered an array with $a=d=0.1\lambda_L$. (a) X gate applied on impurity 1, which results in a transfer from the excited to the ground state of that impurity. For a strong drive, the gate time is $\tau=\pi/2\Omega_{\alpha} \ll \gamma_I^{-1}$. Here, we use $\Omega_{\alpha}=10^{4} \gamma_{I}$. (b) Z gate applied to impurity 1, which results in a phase change of $\pi$ with respect to the other impurities of the system (red trace and left axis). During the process, the populations are left unchanged. (c) Fidelity of single-impurity X and Z gates, after 700 sequential applications. We have considered an array with $a=d=0.1\lambda_L$. (d) Circuit diagram of $\sqrt{\text{iSWAP}}$ gate, which results in excitation transfer from impurity 1 to impurity 2. (e) Error $\epsilon = 1 -\mathcal{F}$ after applying 100 iSWAP gates as a function of the distance $d$ between the impurities. The different traces correspond to varying lattice spacings $a$. In all subfigures, we consider $\gamma_I = 10^{-4} \gamma_L$. }
\label{fig:iswap}
\end{figure}

First, \new{we devise a gate} that allows any desired rotation around the X-axis of the single-qubit Bloch sphere. We perform the $X_R$ rotation on impurity $\alpha$ by applying a strong resonant pulse $\Omega_{\alpha}$ on it. This results in Rabi oscillations between the ground and excited state, given by the $X_R (\phi)$ rotation matrix 
$$
X_R(\phi) = \begin{bmatrix}
cos(\phi/2) &\; isin(\phi/2) \\
isin(\phi/2) &\; cos(\phi/2)
\end{bmatrix},
$$ 
where $\phi = 2\Omega_{\alpha}\tau$. A notable instance of this class of gates is the X-gate, also known as the NOT gate or bit-flip gate, which is equivalent to a $\pi$ rotation and flips the state of a qubit from $\left|g\right\rangle$ to $\left|e_{\alpha}\right\rangle$, and vice versa. Applying the pulse for $\tau=\pi/2\Omega_{\alpha}$, we obtain an iX gate, which is equivalent to X gate up to a global phase. In \fref{fig:iswap}(a), we plot the population of the ground and excited state of an impurity initially in the excited state after applying $X_R$ for different values of $t$ (or analogously $\phi$). 

Since $ \left|\Omega_{\alpha}  \right | \gg \left|\Phi+\phi^*  \right | \sim \gamma_I$, the time taken to \new{simulate} the X gate is much shorter than the characteristic timescale of the array-mediated impurity-impurity interactions ($ \sim \gamma_I$), which have a negligible effect during the single-qubit operation. \new{Consequently, we achieve a remarkably high fidelity exceeding 99\% when simulating the X gate, even with a circuit depth of 700, as exemplified in \fref{fig:iswap} (c).} 

\subsection{Z rotation}

The $Z_R$ rotation introduces a relative phase $\phi$ between ground and excited states of a single qubit or impurity, and is described by the matrix
\begin{equation}
\label{eq: gate_Z}
    Z_R (\phi) = \begin{bmatrix}
e^{-i \phi} & 0 \\
0 & 1
\end{bmatrix}.
\end{equation}

We \new{devise and simulate} the $Z_R$ gate in our system by applying a far-detuned driving field between the excited state and the HMS of a given impurity, such that $\delta_{R} \gg \Omega_{f}$. If all other drives are zero, the amplitudes in these two states are
\begin{subequations}
    \begin{align}
        \dot{c_{\alpha}}(t)&=- \frac{\Gamma_\mathrm{eff}}{2} c_{\alpha}(t)+i\Omega_{f \alpha}f_{\alpha}(t)\\\dot{f_{\alpha}}(t)&=i\left(\delta_{R}+ \frac{i}{2}\gamma_{R}\right)f_{\alpha}(t)+i\Omega_{f \alpha}^*c_{\alpha}(t)
    \end{align}
\end{subequations}
Since $\delta_{R} \gg \Omega_{f \alpha}, \gamma_I$, the fast-evolving HMS state can be adiabatically eliminated by setting $\dot{f}_{\alpha}\rightarrow 0$. We then obtain 
\begin{align}
    \dot{c_{\alpha}}(t)=\left(-\frac{\Gamma_\mathrm{eff}}{2}- i \frac{|\Omega_{f \alpha}|^2}{\delta _{R}}\right)c_{\alpha}(t).
\end{align}
While the HMS state is never populated due to the off-resonance, it introduces a stark shift $\Omega_{z}=|\Omega_{f \alpha}|^2/\delta _{R}$ to the excited state $\left|c_{\alpha}\right\rangle$. As a result, the excited state rotates at a faster frequency than the ground state and acquires a phase $\phi = \Omega_z \tau$ after a time $\tau$, thereby \new{simulating} the $Z_R$ gate given by Eq.~(\ref{eq: gate_Z}).
Setting $\tau=\pi/\Omega_{z}$, we can \new{devise} a Z gate, which corresponds to a $\pi$ rotation around the $z$-axis of the Bloch sphere [\fref{fig:iswap}(b)]. Finally, it is worth noting that $\Omega_{z} \sim \gamma_L \gg \gamma_{I}$. Consequently, the time required to simulate the $Z$ gate is once again significantly shorter than the characteristic timescale of light-induced impurity-impurity interactions, \new{all while maintaining a fidelity well above 99\% for a circuit depth of 700, as illustrated in \fref{fig:iswap} (c).}

\subsection{$\sqrt{\text{iSWAP}}$ Gate}
$\sqrt{\text{iSWAP}}$ is a two-qubit operation that results in universal for quantum computation when used together with the set of single-qubit rotations (see \fref{fig:iswap}(d)) \cite{schuch2003natural}. This two-qubit gate is described by the unitary matrix
    \begin{subequations}
    \label{eq: iswap}
    \begin{align}
     \sqrt{\text{iSWAP}}=\begin{bmatrix}
       1&  0&  0& 0\\ 
       0&  \frac{1}{\sqrt{2}}& \frac{i}{\sqrt{2}} &0 \\ 
       0&  \frac{i}{\sqrt{2}}& \frac{1}{\sqrt{2}} & 0\\ 
       0&  0&  0& 1
    \end{bmatrix}. 
      \end{align}
    \end{subequations}
For instance, applying an $\sqrt{\text{iSWAP}}$ gate on the state $\left |e_{1},g_{2}\right \rangle$, produces the state $\frac{1}{\sqrt{2}}\left|e_{1},g_{2}\right \rangle+\frac{i}{\sqrt{2}}\left | g_{1},e_{2}\right \rangle$.
    
We \new{devise and simulate} the $\sqrt{\text{iSWAP}}$ gate between two impurities of our system by setting all drives to zero, which ensures that the excitation number in the system is conserved. Then, the equations of motion for the system with two impurities is
    \begin{subequations}
    \label{eq: iswap_eom}
    \begin{align}
        \dot{g}(t) =& 0, \\
        \dot{c}_{1}(t)=&-\frac{\Gamma_{eff}}{2}c_{1}(t)-i[\Phi_{12} +\phi_{12}]c_{2}(t), \\
        \dot{c}_{2}(t)=&-\frac{\Gamma_{eff}}{2}c_{2}(t)-i[\Phi_{21} +\phi_{21}]c_{1}(t), \\
        \dot{C}_{12} =& -\Gamma_{eff} C_{12}(t),
    \end{align}
    \end{subequations}
\new{where $g$ denotes the amplitude of the total ground state, $c_1$ and $c_2$ the amplitudes of the states with impurity 1 or 2 populated, and $C_{12}$ the amplitude of the state where both impurities are excited}. Equation~(\ref{eq: iswap_eom}) result in the $\sqrt{\text{iSWAP}}$ gate given by Eq.~(\ref{eq: iswap}) after evolving for a time $t=\frac{\pi}{4(\Phi +\phi^*)}$ [see~\fref{fig:iswap}(d)]. Since both $\Phi_{12}$ and $\phi^*_{12}$ are of the order of $\gamma_I$, the time required to perform this gate is $\propto \gamma_I^{-1}$. That is, the $\sqrt{\text{iSWAP}}$ gate is much slower than the  single qubit gates.  

We can further define the fidelity of the iSWAP operation as the overlap between a target state $|\psi_t\rangle$ and the one obtained by solving the full dynamics of the system $|\psi_d\rangle$, $\mathcal{F} = |\langle \psi_t | \psi_d \rangle|^2$. In Fig.~\ref{fig:iswap}(e), we show the error $\epsilon = 1 - \mathcal{F}$ after 100 consecutive operations of the iSWAP gate as a function of the distance between the impurities $d$ for different lattice spacing $a$. In general, $\epsilon$ increases with $a$ and $d$. If the impurities are in neighboring plaquettes ($d=a$), we can perform $100$ iSWAP operations with small errors rates of about $\epsilon \sim 10^{-3}$ and $\epsilon \sim 10^{-4}$ for $a=0.1\lambda_L$ and $a=0.05\lambda_L$, respectively. While this error remains relatively small for $a=0.05 \lambda_L$ and $d=4a$, it approaches $\epsilon \rightarrow 1$ if the lattice spacing is increased to $a=0.1 \lambda_L$. This sets the maximum distance between impurities for which the iSWAP gate can be applied with high fidelities, and therefore sets the fundamental limit of this platform. 
\new{
The high fidelity obtained for closely spaced impurities is comparable with that observed in other quantum platforms. For instance, a single $\sqrt{\text{iSWAP}}$ gate using superconducting qubits typically exhibits a fidelity of 98.2\% ~\cite{Kjaergaard2020}. Meanwhile, ion-trap-based systems perform a similar iSWAP gate with a fidelity of 97.7\% ~\cite{Drmota2023}. Moreover, we also show in Appendix 
~\ref{appdx: fidelity atomic motion} that the fidelity of our single and two-qubit gates remains intact in the presence of random fluctuations or atomic motion in our lattice array which may take place due to experimental errors. }

\subsection{Decoupling Impurities}

To \new{devise and simulate} any arbitrary quantum circuit, we need to control the interactions between any two qubits or impurities during the whole process. In the platform at hand, however, the $\sqrt{\text{iSWAP}}$ operation continuously acts between all impurities present and may introduce significant errors when performing the slow two-qubit $\sqrt{\text{iSWAP}}$ gate. To avoid this, we require a method that decouples all qubits that are not involved in a specific operation. In what follows, we describe multiple techniques to decouple impurities that are in an arbitrary superposition of the ground and excited state. It is worth noting that such decoupling protocols only need to be implemented when performing an iSWAP gate. The single impurity gates X and Z are significantly faster than the characteristic timescale of the lattice mediated impurity-impurity interactions and are therefore unaffected by them.

\subsubsection{Decoupling impurity in the ground state}
\noindent First, we describe how to decouple an impurity that is in the ground, thus hindering population transfer from other impurities (\ie turning off the iSWAP operation). \\

\noindent \textit{a.) Detuning $|g\rangle \leftrightarrow |e\rangle$ transition}

To decouple impurity $\alpha$, we can apply a detuning to the $|g\rangle \leftrightarrow |e\rangle$ of that impurity larger than the lattice-mediated impurity-impurity interactions, \ie $\delta_\alpha \gg | \Phi+\phi^* | \sim \gamma_I$. This energy shift can be either applied to the ground or excited state, and can be respectively modeled by adding the term $\delta_\alpha s_{\alpha}s_{\alpha}^{\dagger}$ or $\delta_\alpha s_{\alpha}^{\dagger}s_{\alpha}$ in the impurity Hamiltonian given  by Eq.(\ref{eq: Hamiltonian_impurity}). As shown in Appendix~\ref{appendix: EIT}, the excited state can for example be detuned by coupling it to an additional energy level with a strong drive, in a process analogous to electromagnetically induced transparency.

\noindent \textit{b.) Population transfer to another HMS state}

\noindent Alternatively, we can transfer the excitation in the ground state to another auxiliary state that does not decay and does not interact with the remaining impurities via light-induced dipole-dipole interactions. Excitation exchange between both states can simply be performed with a $\pi$-pulse.

\subsubsection{Decoupling impurity in the excited state} \noindent An impurity $\alpha$ in the excited state $|e_\alpha \rangle$ does not decay only at the optimal lattice-impurity detuning $\delta_{LI}$. As opposed to decoupling an impurity in the ground state, we cannot detune the $|g\rangle \leftrightarrow |e\rangle$ transition in this case, since it would result in a quick decay of the excited impurity.

As a result, decoupling an excited impurity requires to transfer the excitation to another metastable state $| r_{\alpha} \rangle$, which again does not decay and does not interact with the other impurities. This is achieved by applying a $\pi$-pulse with strength $\Omega_{f \alpha} \gg | \Phi+\phi^* |$ which promotes $\left | e_{\alpha} \right \rangle \rightarrow \left | r_{\alpha} \right \rangle$. \\

\subsubsection{Decoupling impurity in an arbitrary state} \noindent

If impurity $\alpha$ is in an arbitrary superposition of ground and excited state, we have to combine the methods described above to decouple it. We will do so by first transferring the excited part of the wave function to a metastable state and then detuning the $|g\rangle \leftrightarrow |e\rangle$ transition. Note that any extra phase acquired by the impurity during this process can be simply compensated by applying a local Z-gate.

\subsubsection{\new{Addressing a single site with subwavelength spacing}}

\new{Both detuning the $|g\rangle \leftrightarrow |e\rangle$ transition and transferring population to the HMS state require the selective control of impurity atoms. While individually addressing tightly spaced atoms may pose an experimental challenge, it has been realized experimentally for quite some time \cite{Weitenberg2011}. Such approaches employ high-aperture microscopic objectives to focus off-resonant laser beams on single lattice sites, detuning only the single selected atom into resonance with a more broadly applied driving beam. Remarkably, not only is this method suitable for atom arrays with lattice spacings on the order of an optical wavelength, the full width at half maximum of the focused detuning beam is narrow enough that nearest neighbor atoms experience but $10\%$ of its peak intensity, while its position is accurate enough to position it within a tenth of a lattice spacing.} 

\section{Operations}
\label{section:operations}
The set of gates defined above is universal for quantum computation, and can be therefore used to construct any quantum operation or circuit \cite{schuch2003natural}. As an example, we here \new{theoretically and numerically} demonstrate how to generate two- and three-qubit entangled states.

\subsection{Entangling two impurities: Bell state}
\label{subsection: BellState}
    \begin{figure}
    \centering
    \includegraphics[height=11cm,width=7.5cm]{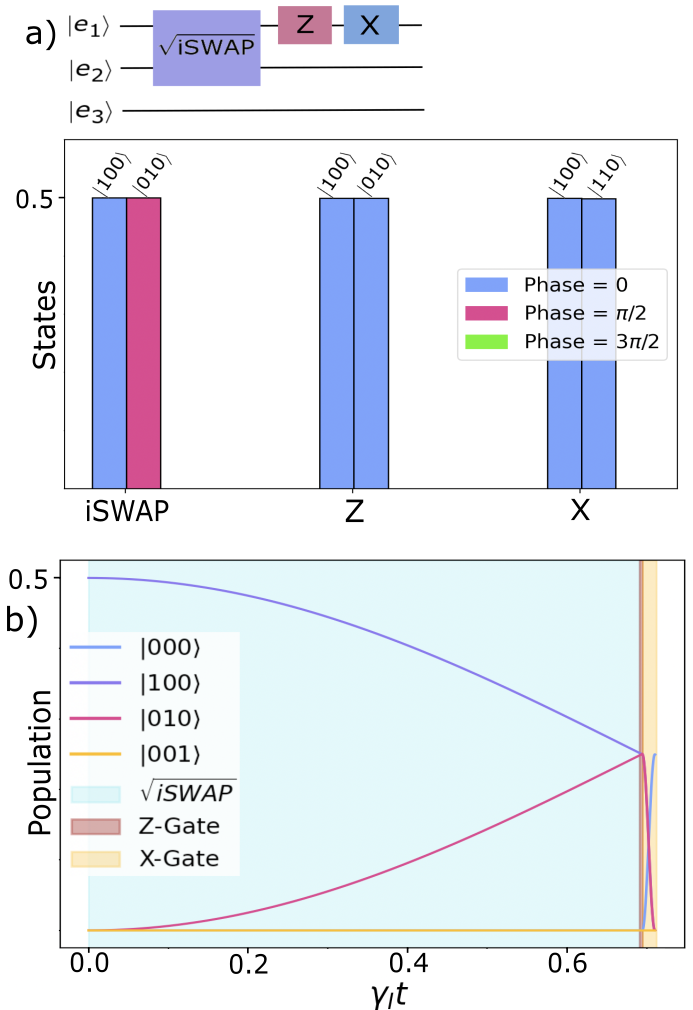}
   \caption{(a) Decomposition of Bell state, $\vert\Phi^{+}\rangle = \frac{1}{\sqrt{2}}(\vert00\rangle + \vert11\rangle)$, using the native gate set of our system, and state of the three-impurity system after each gate. The third impurity is detuned and remains always in the ground state. (b) Temporal evolution of the state of the impurities throughout the protocol (line plot). The different background colors correspond to the different gates applied. We consider $\gamma_I = 10^{-3} \gamma_L$, $d=a=0.1\lambda_L$. Additionally, we use $\delta_{R2} = 200 \gamma_L$ and $\Omega_{f2} = \gamma_L$ when implementing the Z-gate on impurity 2, and  $\Omega_{2} = \gamma_L$ for the X-gate on impurity 2. }
   \label{fig:bell}
    \end{figure}

The Bell states, $\vert\Phi^{\pm}\rangle = \frac{1}{\sqrt{2}}(\vert00\rangle \pm \vert11\rangle)$ and $\vert\Psi^{+}\rangle = \frac{1}{\sqrt{2}}(\vert10\rangle \pm \vert01\rangle)$, are four maximally entangled two-qubit states. That is, measuring of one of the qubits immediately collapses the wavefunction of the other qubit, regardless of the physical distance between them \cite{horodecki2009quantum}.

Figure.~\ref{fig:bell}(a) displays the decomposition of the Bell state $\vert\Phi^{+}\rangle$ using the gate set available in our system. Assuming that only one of the two impurities is initially excited, one needs to applies a $\sqrt{\text{iSWAP}}$ gate on both impurities, followed by a $\frac{\pi}{2}$-rotation around the Z-axis and an X gate on the second qubit. If the system contains extra impurities, we simply need to decouple them by applying a large detuning to render them out of resonance. For the three-impurity system shown in Fig.~\ref{fig:bell}, the prepared state is thus $\vert\Phi^{+}\rangle \otimes |g\rangle$. As shown in Fig.~\ref{fig:bell}(b), the Bell state can be prepared with high fidelity ($>0.999$) for $a= 0.1\gamma_{I}$. This error originates predominantly from the $\sqrt{\text{iSWAP}}$ and follows a similar trend as shown in Fig.~\ref{fig:iswap}(e). Finally, it is worth noting that the remaining Bell states can be obtained by applying suitable combinations of a $\frac{\pi}{2}$-rotations around the Z axis and an X gate on the second qubit.

\subsection{Entangling three impurities: GHZ state}

\begin{figure}
    \centering
    \includegraphics[height=10.5cm,width=7.6cm]{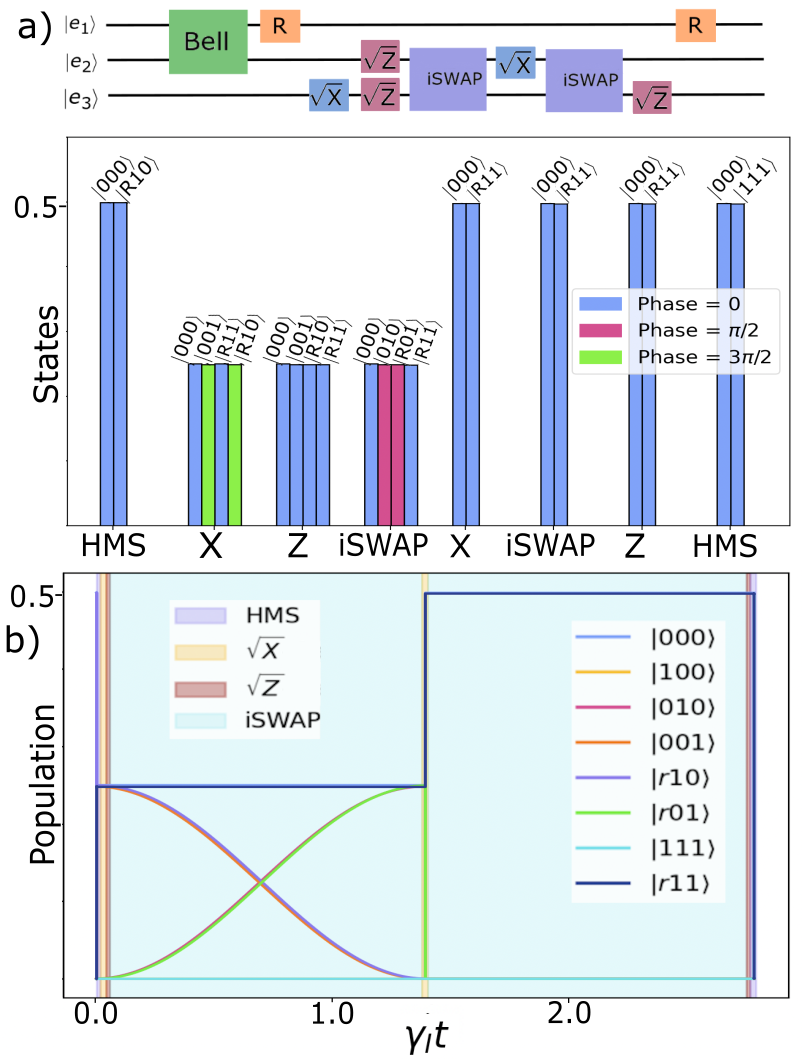}
   \caption{(a) Sequence of gates used to create a three-qubit GHZ state using the native gate set of our system, as well as the intermediate quantum states attained after each gate. (b) Temporal evolution of the state of the impurities throughout the protocol (line plot). Again, the different background colors correspond to the different gates applied. Same parameters as in Fig.~\ref{fig:bell}. }
   \label{fig:ghz}
    \end{figure}

We further \new{devise} the implementation of the entangled three-qubit Greenberger-Horne-Zeilinger (GHZ) state, which is defined as $\vert \text{GHZ} \rangle = \frac{1}{\sqrt{2}}(\vert 000 \rangle + \vert 111 \rangle)$ \cite{bouwmeester1999observation} and has potential applications in various areas such as quantum error correction, quantum teleportation, and quantum cryptography ~\cite{gao2010cryptanalysis, moreno2018using}. The three-qubit GHZ state can be obtained from the Bell state $\vert\Phi^{+}\rangle \otimes |g \rangle$ prepared in Section~\ref{subsection: BellState} by applying a CNOT gate with the second and third qubits as the control and target qubits, respectively. The CNOT gate can be efficiently decomposed into our system's gate set, as shown in the ~\fref{fig:ghz}(a). Finally, we illustrate the evolution of the state of the impurities during the protocol in ~\fref{fig:ghz}(b), where the initial state is taken to be $\vert\Phi^{+}\rangle \otimes |g \rangle$.

\section{Conclusion}
\label{section:conclusion}

We have theoritically and numerically demonstrated that the universal quantum gate set composed of the two-qubit $\sqrt{\text{iSWAP}}$ gate and the single-qubit X and Z rotations can be readily devised and simulated on impurity emitters embedded in atomic lattices with subwavelength spacing, and illustrate their operation by preparing two- and three-qubit entangled states. We have further show that this platform allows for very high gate fidelities, provided that the impurities are sufficiently close to one another. In real experimental implementations, however, additional errors may arise from various sources such as decoherence, environmental noise, and position or frequency disorder of the emitters.
This work paves the way towards controlling and engineering long-range interactions between emitters \cite{Rubiesbigorda2022,Zanner2022,reitz2022cooperative,Tabares2023, Periwal2021,Kannan2023,Mirhosseini2019}, and demonstrates the potential of using subwavelength arrays for quantum simulation \cite{Lloyd1996} and computation \cite{nielsen2001quantum}, as well as for quantum information storage and processing \cite{northup2014quantum,wehner2018quantum}. In future works, more sophisticated protocols such as optimization \cite{Farhi2014,patti2022variational} and error-correction \cite{knill2000theory} could also be explored on this platform.

This theoretical proposal may be implemented using ultracold atoms trapped using optical lattices \cite{anderson2020realization, rui2020subradiant}, optical tweezers \cite{norcia2018microscopic} or metasurface holographic optical traps \cite{huang2022metasurface}. \new{Atomic lattices are a particularly intriguing platform, due to their ability to interface with free space light \cite{Shahmoon2017}, their long coherence times \cite{Patti2021, Masson2020}, and the resulting high fidelities of gate operations (see ~\fref{fig:iswap}). Among such atomic platforms for quantum information, strontium has become a popular proposal, due to its readily driven $\lambda_s=2.6\mu$m transition $^3 P_0 \rightarrow ^3 D_1$, which can be paired with a trapping, magic wavelength of $a_s = \lambda_s / 16.3$ \cite{Janos_topo,Olmos2013}. Likewise, ytterbium's telecom transitions and optical trapping wavelengths render it a promising species \cite{Covey2019}. As discussed in Sec.\ \ref{section:model}, to ensure the feasibility of our impurity vs array atom criteria, we only require that the two atom varieties demonstrate (i) similar transition wavelengths, e.g.,  $|\omega_\text{I} - \omega_\text{L}| \ll \omega_\text{I}$, $\omega_\text{L}$, and (ii) that the decay rate of the impurity transition is much slower than that of the lattice atom, e.g., $\gamma_\text{I} \ll \gamma_\text{L}$. These properties can be furnished by distinct isotopes of a given atomic species, such as the strontium isotopes $^{87}$Sr and $^{88}$Sr \cite{Stellmer2014,Masson2020,Patti2021}, as well as by simply choosing a species that satisfies the latter criterion and ensuring the former by shifting the transition frequency of either the lattice or impurity atoms via an AC stark shift, such as with an off-resonant beam or optical tweezers. Dual-element arrays, which contain two entirely different atomic species, could also be considered and have recently been demonstrated experimentally \cite{Singh2022}.}

\new{A central experimental difficulty of the proposed protocol in atomic lattices lies in individually addressing each impurity emitter. This feature, which is needed to perform the single-qubit gates and to turn on and off the two-qubit iSWAP gate, could utilize a narrow focusing or localization of the driving fields to focal points smaller than an optical transition wavelength. As achieving such culminated beam focus can be challenging, the high-aperature microscopic objective-focused off-resonant laser technique of \cite{Weitenberg2011}, which selectively tunes single atoms in optical lattices into focus with a more ubiquitous transition drive, represents a promising avenue.}

\new{Even in the event that single-site driving fields prove too challenging in free space platforms such as atomic arrays, this difficulty could be circumvented by extending this proposal to other types of baths or structures capable of mediating light-induced interactions. For instance, one-dimensional waveguides have long demonstrated the ability to strongly couple to target atoms with band edge scattering behavior very similar to that furnished by the lattice atoms of this work \cite{Chang2013,Goban2014,Hood2016}. More recently, analogous systems using superconducting qubits \cite{Mirhosseini2019,Kannan2023} and even color centers in silicon \cite{Prabhu2023} have been investigated, opening the possibility of carrying out our proposal in the microwave regime. Likewise, strong, engineerable atomic interactions can be obtained with cavities \cite{Periwal2021}. As a crucial feature, the above discussed platforms provide controllable, strong couplings between emitters that may be separated by distances much larger than the resonant wavelength \cite{Masson2020,Zanner2022,Tabares2023}. Not only do such coupling distances enable individual addressing of each impurity, they facilitate the design of more versatile quantum circuits involving gates between more distant emitters.}  

\begin{acknowledgments}

O.R.B. acknowledges support from Fundación Mauricio y Carlota Botton and from Fundació Bancaria “la Caixa” (LCF/BQ/AA18/11680093). SFY would like to acknowledge funding from NSF through the CUA PFC, PHY-2207972  and the QSense QLCI as well as thank the AFOSR through FA9550-19-1-0233.

\end{acknowledgments}


\bibliographystyle{apsrev4-1-title}
\bibliography{reference_gates}
\pagebreak

\appendix
\section{Equations of Motion}
\label{appendix:EOMS}

The wavefunction of the full system (impurities and lattice atoms) containing at most two excitations is given as,

\begin{widetext}
\begin{align}
\label{eqn:lattice_EOMS}
    \left|\Psi_{1}(t)\right \rangle &= \left|\Psi(t)\right \rangle \otimes |G\rangle+ \sum_{i}^{N_{L}}b_{i}(t)e^{-iw_{I}t}\left|E_{i}, g\right \rangle +\sum_{i}^{N_{L}}\sum_{\alpha}^{N_{I}}v_{i\alpha}(t)e^{-2iw_{I}t}\left|E_{i}, e_{\alpha}\right \rangle +\sum_{i}^{N_{L}}\sum_{\alpha}^{N_{I}}z_{i\alpha}(t)e^{-i(2w_{I}+w_{f})t}\left|E_{i}, r_{\alpha}\right \rangle, 
\end{align}
\end{widetext}
where $|\Psi(t) \rangle$ corresponds to the wavefunction defined in Eq.~(\ref{eqn:6}), and $|E_i,e_\alpha\rangle$, for example, denotes the state where both impurity $\alpha$ and the lattice atom $i$ are excited. Note that we neglect the contribution from the states $|E_i,E_j\rangle$ where two lattice atoms are simultaneously excited. As shown in Appendix~\ref{appendix: multipleexcitationslattice}, the population in the states $|E_i,E_j\rangle$ is strongly suppressed and can therefore be safely neglected.

The dynamics of the system are obtained by solving the Schrödinger equation using the wavefunction in~\eref{eqn:lattice_EOMS},

\begin{widetext}
\begin{subequations}
\label{eqn:EOMs}
\begin{align}
    \dot{a}&=\sum_{\alpha}^{N_{I}}i\Omega_{\alpha }\tilde{c}_{\alpha}(t),\\
    \dot{b}_{i}(t)&= i(\delta_{LI}+\frac{i}{2}\gamma_{L})b_{i}(t)-i\sum_{j\neq i}^{N_{L}} (J_{ij}-\frac{i}{2}\Gamma_{ij})b_{j}(t)-i\sqrt{\frac{\gamma_{I}}{\gamma_{L}}}\sum_{\alpha}^{N_{I}}(J_{i\alpha}-\frac{i}{2}\Gamma _{i\alpha})c_{\alpha}(t)+i\sum_{\alpha}^{N_{I}}\Omega_{\alpha}v_{i\alpha}, \\
    \dot{c}_{\alpha}(t)&=  -\frac{\gamma _{I}}{2}c_{\alpha}(t)-i\sqrt{\frac{\gamma _{I}}{\gamma_{I}}}\sum_{i}^{N_{L}}(J_{\alpha i}-\frac{i}{2}\Gamma _{\alpha i})b_{i}(t)-i\frac{\gamma _{I}}{\gamma _{L}}\sum_{\beta\neq \alpha}^{N_{I}}(J_{\alpha\beta}-\frac{i}{2}\Gamma_{\alpha\beta})c_{\beta}(t)+i\Omega_{\alpha}^{*}a(t)+i\Omega _{f_{\alpha}}f_{\alpha}(t)\\&\nonumber+i\sum_{\beta\neq \alpha}^{N_{I}}\Omega_{\beta}^{*}C_{\alpha\beta}(t),\\
    \dot{f}_{\alpha}(t)&= i(\delta_{R}+\frac{i}{2}\gamma _{R})f_{\alpha}(t) +i\sum_{\beta\neq \alpha}^{N_{I}}\Omega_{\beta}F_{\alpha\beta(t)} +\Omega_{f \alpha}^{*}c_{\alpha}(t),\\
    \dot{v}_{i\alpha}(t) &= i(\delta_{LI}+\frac{i}{2}\gamma _{L}+\frac{i}{2}\gamma _{I})v_{i\alpha}(t)-i\sum_{j\neq i}^{N_{L}}(J_{ij}-i\frac{i}{2}\Gamma_{ij})v_{j\alpha}(t)-i\sqrt{\frac{\gamma_{I}}{\gamma_{L}}}\sum_{\beta\neq \alpha}^{N_{I}}(J_{i\beta}-\frac{i}{2}\Gamma _{i\beta})C_{\alpha\beta}(t)\\& \nonumber-i\frac{\gamma_{I}}{\gamma_{l}}\sum_{\beta\neq \alpha}^{N_{I}}(J_{\alpha\beta}-\frac{i}{2}\Gamma_{\alpha\beta})v_{i\beta}(t)+i\Omega_{\alpha}^{*}b_{i}(t)+i\Omega_{f \alpha}z_{i\alpha}(t), \\
    \dot{C}_{\alpha\beta}(t)&=-\gamma_{I} C_{\alpha\beta}(t)-i\sqrt{\frac{\gamma _{I}}{\gamma_{L}}}\sum_{i}^{N_{L}}(J_{\alpha i}-\frac{i}{2}\Gamma_{\alpha i})v_{i\beta}(t)-i\sqrt{\frac{\gamma _{I}}{\gamma_{L}}}\sum_{i}^{N_{L}}(J_{\beta i}-\frac{i}{2}\Gamma_{\beta i})v_{i\alpha}(t)+i\Omega_{\beta}^{*}c_{\alpha}(t)+i\Omega_{\alpha}^{*}c_{\beta}(t)\\&\nonumber+i\Omega_{f \alpha}y_{\beta\alpha}(t)+i\Omega_{f \beta}y_{\alpha\beta}(t),\\
    \dot{z}_{i\alpha}(t)&=i(\delta_{LI}+\delta_{R}+\frac{i}{2}\gamma_{L}+\frac{i}{2}\gamma_{R})z_{i\alpha}(t)-i\sum_{i\neq j}^{N_{L}}(J_{ij}-\frac{i}{2}\Gamma_{ij})z_{j\alpha}(t)-i\sqrt{\frac{\gamma_{I}}{\gamma_{L}}}\sum_{\beta\neq \alpha}^{N_{I}}(J_{i\beta}-\frac{i}{2}\Gamma_{i\beta})y_{\beta\alpha}(t)+i\Omega_{f \alpha}^{*}v_{i\alpha}(t),\\
    \dot{y}_{\alpha \beta}(t)&=  i(\delta_{R}+\frac{i}{2}\gamma_{I}+\frac{i}{2}\gamma_{R})y_{\alpha \beta}(t)-i\sqrt{\frac{\gamma_{I}}{\gamma_{L}}}\sum_{i}^{N_{L}}(J_{\alpha i}-\frac{i}{2}\Gamma_{\alpha i })z_{i\beta}(t)+i\Omega_{f \beta}^{*}C_{\alpha\beta}(t)+i\Omega_{f \alpha}F_{\alpha \beta}(t)+i\Omega_{\alpha}^{*}f_{\beta}(t),\\
    \dot{F}_{\alpha \beta}(t)&= 2i(\delta_{R}+\frac{i}{2}\gamma_{R})F_{\alpha\beta}(t)+i\Omega_{f \alpha}^{*}y_{\alpha\beta}(t)+i\Omega_{f \beta}^{*}y_{\beta\alpha}(t).
\end{align}
\label{eqn:EOMs}
\end{subequations}
\end{widetext}

Considering $\gamma_{L}\gg\gamma_{I}$, the characteristic time scale of the lattice atoms is much faster than that of the impurities. Further assuming $\Omega_{\alpha}\ll \gamma_{L}$, we can adiabatically eliminate the states containing an excitation in the lattice atoms. In other words, the lattice immediately reaches its steady state after any slow variation of the impurities. Defining the vectors $\vec{b} = [b_1, \hdots, b_{N_L}]^T$ and $\vec{v}_\alpha = [v_{1\alpha}, \hdots, v_{N_L \alpha}]^T$, we obtain
\begin{equation}
\label{eq: appendix_adiabelim_b}
    \dot{\vec{b}}=0 \;\;\rightarrow \;\; \vec{b}= \sum_\alpha L^{-1} \left( C_{\alpha}c_{\alpha}(t) - \Omega_\alpha \vec{v}_\alpha \right),
\end{equation}
where $L$ is an $N_L \times N_L$ matrix characterizing the interactions between lattice atoms and $C_\alpha$ is an $N_L \times 1$ vector describing the couplings between impurity $\alpha$ and the lattice, both defined in Eq.~(\ref{eqn:LandC_define}).

Since $\vec{v}_{\alpha}$ and $\vec{z}_{\alpha} = [z_{1\alpha}, \hdots, z_{N_L \alpha}]^T$ also contain factors proportional to $\gamma_{L}$, they can also be adiabatically eliminated. Keeping to terms to leading order in the small parameter $\gamma_I/\gamma_L$, we obtain
\begin{subequations}
\begin{align}
\dot{\vec{v}}_{\alpha}&=0 \;\;\rightarrow \;\; \vec{v}_{\alpha}=L^{-1} C_{\beta}C_{\alpha\beta}(t), \\
\dot{\vec{z}}_{\alpha}&=0 \;\;\rightarrow \;\; \vec{z}_{\alpha}=L^{-1} C_{\beta}y_{\beta\alpha}(t).
\end{align}
\label{eqn:nine}
\end{subequations}

Substituting these values in Eq.~(\ref{eq: appendix_adiabelim_b}), we find that the term proportional to $\vec{v}_\alpha$ is suppressed with respect to that proportional to $c_\alpha$ by a factor $\Omega_\alpha/\gamma_L$, and can thus be neglected. Finally, we substitute these instantaneous steady state amplitudes for $\vec{b}$, $\vec{v}_{\alpha}$ and $\vec{z}_{\alpha}$ in~\eref{eqn:EOMs} and obtain the effective equations of motion for the impurity emitters only, given in~\eref{eqn:reduced_EOMS}. 



Finally, it is worth noting that we work in the regime $\Omega_\alpha \sim \gamma_L$ when implementing the single-qubit gates described in Section~\ref{section:gates}. In that case, the driving strength in the impurity is much larger than the light-mediated lattice-impurity interactions, which can therefore be safely neglected for timescales $\tau \sim \gamma_L^{-1} << \gamma_I^{-1}$. In other words, the impurity and lattice atoms simply decouple in this regime.

\section{Effect of multiple excitations in lattice atoms}
\label{appendix: multipleexcitationslattice}

\begin{figure}
    \centering
    \includegraphics[width = \columnwidth]{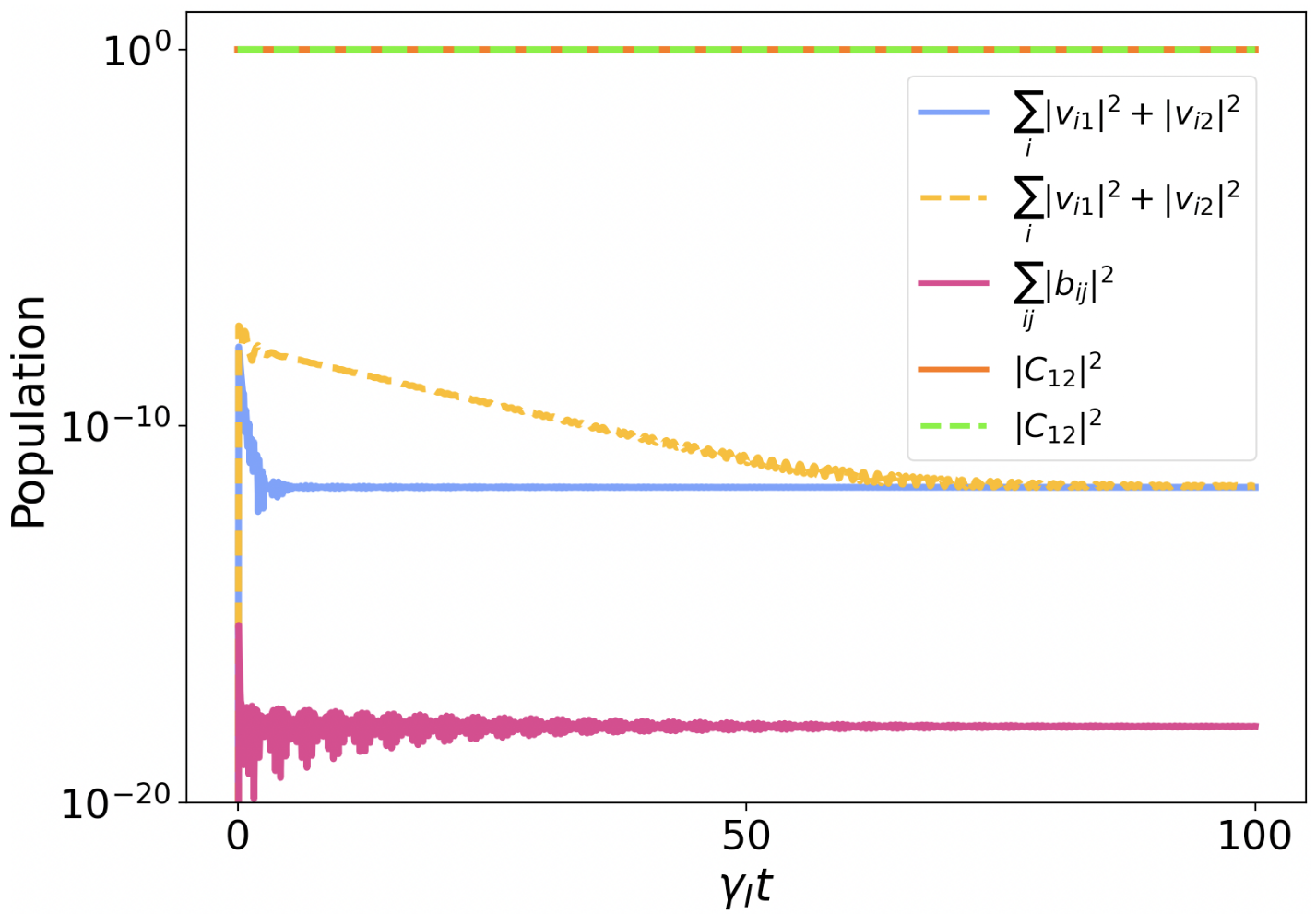}
    \caption{We consider two initially excited impurities embedded in an atomic array. Population for both impurities excited ($|C_{12}|^2$), one impurity and one lattice atom excited ($\sum_{i} |v_{i1}|^2 + |v_{i2}|^2$), and two lattice atoms excited ($\sum_{i,j} |b_{ij}|^2$) as a function of time. The solid lines correspond to the full wavefunction that contains terms corresponding to two lattice atoms excited. The dashed lines correspond to a simulation using the approximate wavefunction that neglects two excitations in the lattice. We consider a $10 \times 10$ lattice with spacing $a=0.1\lambda_L$ and distance between impurities $d=a$.}
   \label{fig:app}
\end{figure}

In Appendix~\ref{appendix:EOMS}, we neglect the contribution from the states containing two excitations in the lattice atoms. Here, we justify this approximation. 

For that, let us first consider a single impurity embedded in an atomic array. If no driving fields are applied, the wavefunction simply reads
\begin{equation}
    \left|\psi'\right\rangle = c_{\alpha}\left|e_{\alpha}\right\rangle + \sum_{i}^{N_{L}}b_{i}\left|E_{i}\right\rangle.
\end{equation}
Adiabatically eliminating the lattice atoms, we obtain $\vec{b} =L^{-1}\hat{C}_{\alpha}c_{\alpha}$. Then,

\begin{equation}
\left \langle \psi'|\psi'\right \rangle =(1+\vec{b}^{\dagger}\vec{b})|c_{\alpha}|^{2}\rightarrow |{c_{\alpha}}^{2}|=\frac{1}{1+\vec{b}^{\dagger}\vec{b}}
\end{equation}
Above the band edge region (that is, if the impurity is off-resonant with all normal modes of the lattice), the impurity population $|{c_{\alpha}}^{2}|$ is the vast majority of the wavefunction and the population in the lattice atoms $b_i$ is hence negligibly small. Performing a similar analysis for a two-impurity system, we can conclude that the terms $b_{ij}$ containing both excitations in the lattice are even further suppressed.

We can numerically confirm this intuition by simulating the dynamics of the full system. For that, we write down the full wavefunction containing up to two excitations,
\begin{align}
    \left|\Psi_2(t)\right\rangle= |\Psi_1(t) \rangle + \sum_{i}^{N_{L}}\sum _{j\neq i}^{N_{I}}b_{ij}(t)e^{-2i\omega_{I} t} \left | E_{i},E_{j},g\right \rangle,
\end{align}
where $|\Psi_1(t) \rangle$ is given in Eq.~(\ref{eqn:lattice_EOMS}). Applying the Schrodinger equation, we find the equations of motion for the terms $b_{ij}$,
\begin{widetext}
    \begin{align}
    \dot{b}_{ij}(t)= 2i(\delta_{LI}+\frac{i}{2}\gamma_{L})b_{ij}(t)-i\sum_{i\neq j}^{N_{L}}[({J_{ij}-\frac{i}{2}\Gamma_{ij}})b_{ij}(t)+({J_{ji}-\frac{i}{2}\Gamma_{ji}})b_{ji}(t)]-i\sqrt{\frac{\gamma_{I}}{\gamma_{L}}}\sum_{i\neq j}^{N_{L}}\sum_{\beta \neq \alpha}^{N_{I}}(J_{i\alpha}-\frac{i}{2}\Gamma_{i \alpha})v_{i \alpha}(t).
   \end{align}
\end{widetext}

In~\fref{fig:app}, we compare the dynamics resulting from the full set of equations containing two excitations in the lattice (solid lines) with the approximated version used in Appendix~\ref{appendix:EOMS} that neglects the terms $b_{ij}$ (dashed lines). The population in the impurity emitters, as well as in the terms containing one excitation in the lattice and one in an impurity ($\sum_{i,\alpha} |v_{i \alpha}|^2$), are approximately equal in both cases. As expected, the population $\sum_{i,j} |b_{i j}|^2$ is heavily suppressed and can thus be safely neglected. It is worth noting that this approximation largely reduces the computational power required to simulate the dynamics of the system.

\section{Equations of motion for three excitations}
To implement the entangled Bell and GHZ states in Section~\ref{section:operations}, we need to consider states containing up to three excitations. Then, the total wavefunction takes the form
\begin{widetext}
        \begin{align}
            \left|\Psi_{2}(t)\right \rangle &= \left|\Psi_{1}(t)\right \rangle+
            \sum ^{N_{I}}_{\alpha,\beta,\epsilon \neq \alpha,\beta }C_{\alpha\beta\epsilon}(t)e^{-3iw_{I}t}\left|G,e_{\alpha},e_{\beta},e_{\epsilon}\right\rangle + \sum ^{N_{I}}_{\alpha,\beta,\epsilon \neq \alpha,\beta }F_{\alpha\beta\epsilon}(t)e^{-i3(w_{I}+w_{R})t}\left|G,r_{\alpha},r_{\beta},r_{\epsilon}\right\rangle  \nonumber \\ & \nonumber+ \sum ^{N_{L}}_{i}\sum ^{N_{I}}_{\alpha,\beta\neq\alpha}v_{i\alpha\beta}(t)e^{-i3w_{I}t}\left|E_{i},e_{\alpha},e_{\beta}\right\rangle+\sum ^{N_{L}}_{i}\sum ^{N_{I}}_{\alpha,\beta\neq\alpha}z_{i\alpha\beta}(t)e^{-i(3w_{I}+2w_{R})t}\left|E_{i},r_{\alpha},r_{\beta}\right\rangle \\ & \nonumber+\sum ^{N_{I}}_{\alpha,\beta,\epsilon\neq\alpha,\beta}E_{\alpha\beta\epsilon}(t)e^{-i(3w_{I}+w_{R})t}\left |r_{\alpha},e_{\beta},e_{\epsilon}\right\rangle +\sum_{\alpha,\beta,\epsilon\neq\alpha,\beta}G_{\alpha\beta\epsilon}(t)e^{-i(3w_{I}+2w_{R})t}\left|e_{\alpha},r_{\beta},r_{\epsilon}\right\rangle +\\ &  \sum_{i}^{N_{L}}\sum_{\alpha,\beta\neq\alpha}^{N_{I}}R_{i\alpha\beta}(t)e^{-i(3w_{I}+w_{R})t}\left|E_{i},r_{\alpha},e_{\beta}\right\rangle. 
        \end{align}
\end{widetext}
Using the Hamiltonian in Eq.~(\ref{eqn:EOMs}) and the Schrödinger equation, we obtain the equations of motion describing the full system. For concision, we only provide the equations that differ from those in \eref{eqn:EOMs},
\begin{widetext}
    \begin{subequations}
        \begin{align}
        \dot{C}_{\alpha\beta}(t)=& \cdots+i\sum_{\epsilon\neq \alpha,\beta}^{N_{I}}\Omega_{\epsilon}C_{\alpha\beta\epsilon}(t),\\
        \dot{z}_{i\alpha}(t)=&\cdots +i\sum_{\beta\neq\alpha}^{N_{I}}\Omega_{\beta}R_{i\alpha\beta}(t),\\
        \dot{y}_{\alpha\beta}(t)=&\cdots +i\sum_{\epsilon\neq\alpha,\beta}^{N_{I}}\Omega_{\epsilon}E_{\alpha\beta\epsilon}(t),\\
        \dot{F}_{\alpha\beta}(t)=&\cdots +i\sum_{\epsilon\neq\alpha,\beta}^{N_{I}}\Omega_{\epsilon}G_{\epsilon\alpha\beta}(t),\\
        \dot{C}_{\alpha\beta\epsilon}(t)=&-\frac{3}{2}\gamma_{I}C_{\alpha\beta\epsilon}(t)+\sum^{N_{L}}_{i}\sum _{\alpha}^{N_{I}}(J_{\alpha i}-\frac{i}{2}\Gamma_{\alpha i})v_{i\beta\epsilon}(t)+i\sum_{\alpha}^{N_{I}}\Omega_{\alpha}^{*}C_{\beta\epsilon}(t)+i\sum_{\alpha}^{N_{L}}\Omega_{f \alpha}E_{\alpha\beta\epsilon}(t),\\
        \dot{R}_{\alpha\beta\epsilon}(t)=& \;3i(\delta_{R}+\frac{i}{2}\gamma _{R})R_{\alpha\beta\epsilon}(t)+i\sum_{\alpha}^{N_{L}}\Omega_{f \alpha}^{*}G_{\alpha\beta\epsilon}(t),\\
        \dot{v}_{i\alpha\beta}(t)=& \;i(\delta_{LI}+\frac{i}{2}\gamma_{L}+i\gamma_{I})v_{i\alpha\beta}(t)-i\frac{\gamma_{I}}{\gamma_{L}}\sum^{N_{I}}_{\alpha,\epsilon\neq\alpha,\beta}(J_{\alpha\epsilon}-\frac{i}{2}\Gamma_{\alpha\epsilon})v_{i\epsilon\beta}(t)-i\frac{\gamma_{I}}{\gamma_{L}}\sum_{\epsilon\neq\alpha\beta}^{N_{I}}(J_{i\epsilon}-\frac{i}{2}\Gamma_{i\epsilon})C_{\alpha\beta\epsilon}(t)\\&-i\sum_{j\neq i}^{N_{L}}(J_{ij}-\frac{i}{2}\Gamma_{ij})v_{j\alpha\beta}(t) +i\sum _{\alpha}^{N_{I}}\Omega_{\alpha}^{*}v_{i\alpha}(t)+i\sum_{f_{\alpha}}^{N_{I}}\Omega_{f \alpha}R_{i\alpha\beta}(t),\\
        \dot{z}_{i\alpha\beta}(t)=&i(\delta_{LI}+\frac{i}{2}\gamma_{L}+2\delta_{R}+i\gamma_{R} )z_{i\alpha\beta}(t)-i\frac{\gamma_{I}}{\gamma_{L}}\sum_{\epsilon\neq\alpha\beta}^{N_{I}}(J_{i\epsilon}-\frac{i}{2}\Gamma_{i\epsilon})G_{\epsilon\alpha\beta}(t)-i\sum_{j\neq i}^{N_{L}}(J_{ij}-\frac{i}{2}\Gamma_{ij})z_{j\alpha\beta}(t) +i\sum_{f_{\alpha}}^{N_{I}}\Omega_{f \alpha}^{*}R_{i\beta\alpha}(t),\\
        \dot{E}_{\alpha\beta\epsilon}(t)=&i(\delta_{R}+\frac{i}{2}\gamma_{R}+i\gamma_{I})E_{\alpha\beta\epsilon}(t)-i\sqrt{\frac{\gamma_{I}}{\gamma_{L}}}\sum_{i}^{N_{L}}\sum_{\beta}^{N_{I}}(J_{\beta i}-\frac{i}{2}\Gamma_{\beta i})R_{i\alpha\epsilon}(t)+i\Omega_{f \alpha}^{*}C_{\alpha\beta\epsilon}+i\sum_{\beta\neq\alpha}^{N_{I}}\Omega_{\beta}^{*}y_{\beta\alpha}(t)+i\sum_{\beta\neq\alpha}^{N_{I}} \Omega_{f \beta}G_{\epsilon\beta\alpha}(t),\\
        \dot{G}_{\alpha\beta\epsilon}(t)=&i(2\delta_{R}+{i}\gamma_{R}+\frac{i}{2}\gamma_{I})G_{\alpha\beta\epsilon}(t)-i\sqrt{\frac{\gamma_{I}}{\gamma_{L}}}\sum_{i}^{N_{L}}(J_{\alpha i}-\frac{i}{2}\Gamma_{\alpha i})Z_{i\beta\epsilon}(t)+i\Omega_{\alpha}^{*}F_{\beta\epsilon}+i\sum_{\beta\neq\alpha}^{N_{I}}\Omega_{f \beta}^{*}E_{\epsilon\beta\alpha}(t)+i\Omega_{f \alpha}F_{\alpha\beta\epsilon}(t),\\
        \dot{R}_{i\alpha\beta}(t)=&i(\delta_{LI}+\frac{i}{2}\gamma_{L}+\frac{i}{2}\gamma_{I}+\delta_{R}+\frac{i}{2}\gamma_{R})R_{i\alpha\beta}(t)-i\sqrt{\frac{\gamma_{I}}{\gamma_{L}}}\sum_{j\neq i}^{N_{L}}(J_{ij}-\frac{i}{2}\Gamma_{ij})R_{j\alpha\beta}(t)-i\frac{\gamma_{I}}{\gamma_{L}}\sum _{\epsilon\neq\alpha,\beta}^{N_{I}}(J_{\beta\epsilon}-\frac{i}{2}\Gamma _{\beta\epsilon})R_{i\alpha\epsilon}(t)\\ \nonumber &-i\sqrt{\frac{\gamma_{I}}{\gamma_{L}}}\sum _{\epsilon\neq\alpha,\beta}^{N_{I}}(J_{i\epsilon}-\frac{i}{2}\Gamma _{i\epsilon})E_{\alpha\beta\epsilon}(t)+i\Omega_{f \alpha}^{*}v_{i\alpha\beta}+i\sum_{\beta\neq\alpha}^{N_{I}}\Omega_{\beta}z_{i\alpha}(t)+i\sum_{\beta\neq\alpha}^{N_{I}}\Omega_{f \beta}z_{i\alpha\beta}(t).
        \end{align}
    \end{subequations}
\end{widetext}
Here, the three dots stand for the terms already present in \eref{eqn:EOMs}. Note also that we have negelected all terms containing more than one excitation in the lattice atoms, as justified in Appendix~\ref{appendix: multipleexcitationslattice}.

\section{Decoupling ground state impurities through electromagnetically induced transparency}
\label{appendix: EIT}

To decouple an impurity in the ground state from the rest of the system, we simply need to detune its resonance frequency. This can be achieved by considering impurities with three levels: the ground state $|g_\alpha\rangle$, the dipole-coupled excited state $|e_\alpha\rangle$ and a non-interacting state $|r_\alpha\rangle$. 
For simplicity, let us consider a two-impurity system such that impurity $2$ is in the excited state $|e_2\rangle$ and impurity $1$ is in the ground state $|g_1\rangle$. 
To suppress population transfer from impurity $2$ to $1$, we apply a drive on impurity 1 with strength much larger than the lattice-mediated interactions, \ie $\left |\Omega_{f 1}\right | \gg \left | \Phi_{12}+\phi_{12} \right | \sim \gamma_I$.
The dynamics of the system are then governed by the following Hamiltonian
\begin{widetext}
\begin{align}
\label{eq: app_Hamiltonian_EIT}
    \hat{H} & = -i \frac{\Gamma_{eff}}{2}\big( \left|e_1\right\rangle\left\langle e_1 \right| 
    + \left|e_2\right\rangle\left\langle e_2 \right| \big) +\left(\Phi_{12}+\phi_{12}\right) \big( \left| e_1\right\rangle\left\langle e_2\right| + \left| e_2\right\rangle\left\langle e_1\right| \big) \\ \nonumber
    &-i \frac{\gamma_{R}}{2} \big( \left|r_1\right\rangle\left\langle r_1\right| +\left|r_2\right\rangle\left\langle r_2\right| \big)   -\Omega_{f 1} \big( \left|e_{1}\right\rangle\left\langle r_{1}\right|+ \left|r_{1}\right\rangle\left\langle e_{1}\right| \big),
\end{align}
\end{widetext}
where we have assumed that the drive is on resonance with the $|e_1 \rangle \leftrightarrow |r_1\rangle$ transition, and that its Rabi frequency $\Omega_{f 1}$ is real. Due to the strong drive, the coupling between both impurities can be treated perturbatively. Further assuming that $\Gamma_{eff}, \gamma_R \ll | \Phi_{12}+\phi_{12} | \ll \Omega_{f1}$, we can neglect the decay rates. Then, the eigenstates of impurity $1$ are approximately the dressed states $|\chi_\pm \rangle \approx (|e_{1}\rangle \pm |r_{1}\rangle)/\sqrt{2}$, which are respectively shifted by $\mp \Omega_{f1}$ from the resonance frequency, and the Hamiltonian in Eq.~\ref{eq: app_Hamiltonian_EIT} to lowest order in the perturbation to $\Omega_{f1}$ simply reads
\begin{align}
    \hat{H} & \approx - \Omega_{f 1} |\chi_+ \rangle \langle \chi_+ | + \Omega_{f 1} |\chi_- \rangle \langle \chi_- | \nonumber \\
    &+ \frac{1}{\sqrt{2}}\left(\Phi_{12}+\phi_{12}\right) \sum_{\nu=\pm} \big( |\chi_\nu \rangle \langle e_2 | + |e_2\rangle \langle \chi_\nu |  \big),
\end{align}
where we have omitted the uncoupled state $|r_2\rangle$. That is, the coupling between the excited state of impurity $2$, $|e_2 \rangle$, and the normal modes of impurity $1$, $|\chi_\pm \rangle$ is much smaller than the frequency difference between these states. This strong off-resonance strongly suppresses the population transfer between both emitters, such that impurity $1$ remains in the ground state and is effectively decoupled from the remaining impurities.

We can alternatively understand this phenomenon as an interference effect analogous to electromagnetically induced transparency (EIT). For that, it is enough to note that the state $|e_1\rangle$ is coupled by $\hat{H}$ to the superposition
\begin{equation}
    |B\rangle \propto \Omega_{f1} |r_1\rangle - \left(\Phi_{12}+\phi_{12}\right) |e_2\rangle,
\end{equation}
and is consequently decoupled from the orthogonal state,
\begin{equation}
    |D\rangle \propto \Omega_{f1} |e_2\rangle + \left(\Phi_{12}+\phi_{12}\right) |r_1\rangle.
\end{equation}
That is, the interference between the drive to $|r_1\rangle$ and the light-induced coupling to $|e_2\rangle$ renders $|e_1\rangle$ dark with respect to $|D\rangle$. For $\left |\Omega_{f 1}\right | \gg \left | \Phi_{12}+\phi_{12} \right |$, the dark or uncoupled state takes the form $|D\rangle \approx |e_2\rangle$, and an excitation initially in impurity $2$ is not transfered to impurity $1$. In other words, the $\sqrt{\text{iSWAP}}$ operation $\left|e_{2}\right\rangle\rightarrow\left|e_{1}\right\rangle$ is no longer  permitted.

\section{Impact of atomic motion in lattice on single and two-qubit gate fidelity}
\label{appdx: fidelity atomic motion}
\begin{figure}
\centering
\includegraphics[width=0.5\textwidth]{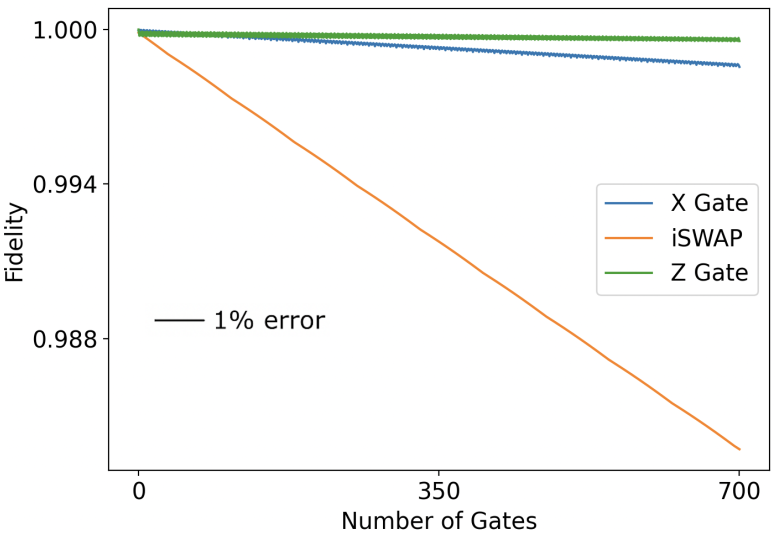}
\caption{Impact of random atomic motion on single and two-qubit gate fidelity by adding Gaussian noise with 1\% standard deviation. We have considered an array with $a=d=0.1\lambda_L$. }
\label{fig:fid_error}
\end{figure}

\new{ We examine the influence of random fluctuations in atomic lattice positions, arising from thermal motion and experimental errors, on the precision of both our single-qubit and two-qubit gate operations. Employing Gaussian noise with a zero mean and standard deviations equivalent to 1\% of the lattice spacing, we perform numerical simulations to capture these atomic displacements. As illustrated in ~\fref{fig:fid_error}, we achieve excellent fidelity for 1\% error, although much more precise lattices are built in practice~\cite{Mcdonald2019}. In fact, even a 1\% error margin overestimates the maximum fluctuations typically observed experimentally~\cite{Mcdonald2019}. These observations underscore the robustness of our single and two-qubit gates. This aligns with prior analytical investigations ~\cite{Shahmoon2017}, where the dispersive shift was found to follow the equation $\left \langle \delta \Delta \right \rangle \approx 4\pi ^2 \left ( \delta_{r}/\lambda \right )^2 \Delta$, with $\delta_{r}$ representing statistically independent and identically-distributed atomic displacement fluctuations, where $\delta_r \ll \Delta$ for realistic atomic placement errors in a subwavelength lattice.}

\section{Heisenberg-Langevin Equation Derivation}
\label{appdx: master eqn}
\new{We address the Fourier Transformed wave equation within polarizable media, employing units where $c = \hbar = 1$. Then the  governing equation can be expressed as follows, 
\begin{equation}
 \vec{\nabla} \times \vec{\nabla} \times \vec{E} - \omega^2 \vec{E} = \frac{\omega^2}{\varepsilon_0} \vec{P}.
\end{equation}
Here, the solution is decomposed into distinct photon polarizations, denoted as $\nu$, and can be expressed through the Green's function $\overline{\overline{G}}(r,\omega)$.
\begin{align}
\vec{E}^{+} (\vec{r},\omega) &= \vec{E}_0^{+}  (\vec{r},\omega) + \vec{E}_p^{+}  (\vec{r},\omega) \\&= \vec{E}_0^{+} (\vec{r},\omega) + \frac{\omega^2}{\varepsilon_0} \int d^{3} \vec{r}' \, \bar{\bar{G}}(\vec{r}-\vec{r}',\omega) \cdot \vec{P}^{+} (\vec{r}',\omega),
\end{align}
where $\vec{E}_0^{+}  (\vec{r},\omega)$ represents the positive homogeneous field solution and the negative frequency counterpart of this field can be obtained via Hermitian conjugate. It's noteworthy that this operator encompasses all electric field components unrelated to polarization or media scattering. In our context, these include contributions from both vacuum and applied driving fields. When considering an atomic lattice defined by positions $\vec{r}_i$ and a single impurity atom situated at $\vec{r}_s$, the expression for the polarization $\vec{P}^{+}$ in the medium can be written as, 
\begin{align}
    \vec{P}^+(\vec{r},\omega) = \sum_\nu \left(\sum_j \left[\vec{d}_{j\nu} \sigma_{j\nu}(\omega) \delta(\vec{r} - \vec{r}_i) \right ]+ \vec{d}_{s\nu} s_\nu(\omega) \delta(\vec{r} - \vec{r}_s) \right).
\end{align}
Here, $\vec{d}_{j\nu}$ and $\vec{d}_{s\nu}$ are the transition dipole moments associated with the atomic lattice positions $\vec{r}_i$ and the impurity atom position $\vec{r}_s$, respectively. The raising operators for lattice atoms and impurity are described as $\sigma^\dagger_{j\nu}(\omega)$ and $s^\dagger_\nu(\omega)$, respectively, and depend on the frequency $\omega$. The solution for the positive frequency component of the field $\vec{E}_p^{+}$ is given as follows,
\begin{widetext}
\begin{align}
    \vec{E}^{+}_p (\vec{r},\omega) &= \frac{\omega^2}{\varepsilon_0} \sum_{\nu} \left (\sum_{j} \left[\bar{\bar{G}}(\vec{r}-\vec{r}_j,\omega) \cdot \vec{d}_{j\nu} \sigma_{j\nu}(\omega)\right] + \bar{\bar{G}}(\vec{r}-\vec{r}_s,\omega) \cdot \vec{d}_{s\nu} s_{\nu} (\omega)\right).
\end{align}
\end{widetext}
At this point, we can take the Markov Approximation by assuming that the reservoir (vacuum) reaches equilibrium much faster than our system (lattice). Then, we can approximate that $\sigma^{\dagger}_{\beta}(\omega), s^{\dagger}_{\beta}(\omega)$ are sharply peaked at their respective resonance frequencies ($\omega_0$ and $\omega_I$). This enables us to approximate the remaining terms of $\vec{E}^{+}_{p}(\vec{r},\omega)$ stationary around this point such that, taking the reverse Fourier Transform back into the time domain we obtain,
\begin{widetext}
\begin{align}
 \vec{E}^{+}_p (\vec{r},t) &=\int_{-\infty}^{\infty} \frac{d\omega}{2\pi} \vec{E}^{+}_p(\vec{r},\omega)e^{-i\omega t} \\ \nonumber  &\approx \frac{\omega^2}{\varepsilon_0}\sum_{\nu} \left ( \sum_{j} \left[\bar{\bar{G}}(\vec{r}-\vec{r}_j,\omega_0) \cdot \vec{d}_{j\nu}\int_{-\infty}^{\infty} \frac{d\omega}{2\pi}\sigma_{j\nu}(\omega) e^{-i\omega t}  \right]+  \bar{\bar{G}}(\vec{r}-\vec{r}_s,\omega_0) \cdot \vec{d}_{s\nu} \int_{-\infty}^{\infty} \frac{d\omega}{2\pi}s_{\nu} (\omega) e^{-i\omega t}\right) \\ \nonumber
    &= \frac{\omega^2}{\varepsilon_0} \sum_{\nu}\left (   \sum_{j} \left[\bar{\bar{G}}(\vec{r}-\vec{r}_j,\omega_0) \cdot \vec{d}_{j\nu} \sigma_{j\nu}(t)\right]+ \bar{\bar{G}}(\vec{r}-\vec{r}_s,\omega_0) \cdot \vec{d}_{s\nu} s_{\nu} (t)\right ).
\end{align}
\end{widetext}
In this expression, we have assumed that $|\omega_I - \omega_0| \ll \omega_I, \omega_0$. The sharp peak assumption allows us to treat \( \sigma_{j\nu} \) and \( \sigma_{s\nu} \) as approximately constant around \( \omega_0 \). The system operator Q can be characterized by its evolution under the Hamiltonian $H=H_{LI}+V_{AF}$ where, 
\begin{align}
H_{LI}&=\omega_0\sum_{ui}\sigma_{iu}^{\dagger}\sigma_{iu} + \omega_{I}\sum_{u}s_{u}^{\dagger}s_{u}. \\
V_{AF}&=-\vec{E} \cdot \hat{d} 
\end{align} where, 
\begin{align}
    \nonumber
    \hat{d}=\sum_{iu}\left [  \vec{d}^{\dagger}_{iu}\sigma_{iu}^{\dagger}(t)+\vec{d}_{iu}\sigma_{iu}(t) \right ]+\vec{d}^{\dagger}_{su}s_{u}^{\dagger}(t)+\vec{d}_{su}s_{u}(t)
\end{align}
As the time-dependent operators in $\hat{d}$ permit us to use Rotating Wave Approximation to isolate slowly oscillating terms, the portion of the interaction due to the positive frequency term is,
\begin{align}
    V_{AF}^{+}&=-\sum_{u}\left [  \sum_{i} \vec{d}^{\dagger}_{iu}\sigma_{iu}^{\dagger}(t)+\vec{d}^{\dagger}_{su}s_{u}^{\dagger}(t)\right ] \cdot (\vec{E}_{0}^{+}(\vec{r},t)\\ \nonumber &+\vec{E}_{p}^{+}(\vec{r},t)).
\end{align}
The Heisenberg-Langevin Equation of motion for Q is,
\begin{equation}
    \dot{Q}=-i[Q,H]=-i[Q,H_{LI}+V_{AF}].
\end{equation}
Here, 
\begin{widetext}
  \begin{align}
  \label{eq: appendix_Heisenberg_HLI}
        -i[Q,H_{LI}]&=-i\omega_0\sum_{u}\left \{ \sum_{i} \left ( [Q,\sigma_{iu}^{\dagger}]\sigma_{iu}+ \sigma_{iu}^{\dagger}[Q,\sigma_{iu}]\right ) +[Q,s_{u}^{\dagger}]s_{u}+s_{u}^{\dagger}[Q,s_{u}]\right \} \\
        -i[Q,V_{AF}]&=i\sum_{u}\left \{ \sum_{i} \left ( [Q,\sigma_{iu}^{\dagger}]\vec{d}^{\dagger}_{iu}\cdot \vec{E}_{0}^{+}(\vec{r}_{i},t) \right )+[Q,s_{u}^{\dagger}]\vec{d}^{\dagger}_{su}\cdot \vec{E}_{0}^{+}(\vec{r}_{s},t)\right \}
        \\ \nonumber &+i\sum_{u}\left \{ \sum_{i}\left ( \vec{E}_{0}^{-}(\vec{r}_{i},t)\cdot \vec{d}_{iu}[Q,\sigma_{iu}] \right )+\vec{E}_{0}^{-}(\vec{r}_{s},t) \cdot \vec{d}_{su}[Q,s_{u}]\right \}
        \\ \nonumber &+i\sum_{u}\left \{ \sum_{i} \left ( [Q,\sigma_{iu}^{\dagger}]\vec{d}^{\dagger}_{iu}\cdot \vec{E}_{p}^{+}(\vec{r}_{i},t) \right )+[Q,s_{u}^{\dagger}]\vec{d}^{\dagger}_{su}\cdot \vec{E}_{p}^{+}(\vec{r}_{s},t)\right \}
        \\ \nonumber &+i\sum_{u}\left \{ \sum_{i}\left ( \vec{E}_{p}^{-}(\vec{r}_{i},t)\cdot \vec{d}_{iu}[Q,\sigma_{iu}] \right )+\vec{E}_{p}^{-}(\vec{r}_{s},t) \cdot \vec{d}_{su}[Q,s_{u}] \right \}
  \end{align}
\end{widetext}
Taking the Hermitian conjugate of $H_{E}^+$ and employing the properties 
that $\bar{\bar{G}}(\vec{r}_i, \vec{r}_j ) = \bar{\bar{G}}(\vec{r}_j, \vec{r}_i)$ and $\bar{\bar{G}}^{\dagger}(\vec{r}_i, \vec{r}_j )=\bar{\bar{G}}^{\ast}(\vec{r}_i, \vec{r}_j)$ on the particular solution part of the expression we get, 
\begin{widetext}
    \begin{align}
    \label{eq: appendix_Langevin_QVAF}
    -i[Q,V_{AF}]&=i\sum_{u}\left \{ \sum_{i} \left ( [Q,\sigma_{iu}^{\dagger}]\vec{d}^{\dagger}_{iu}\cdot \vec{E}_{0}^{+}(\vec{r}_{i},t) \right )+[Q,s_{u}^{\dagger}]\vec{d}^{\dagger}_{su}\cdot \vec{E}_{0}^{+}(\vec{r}_{s},t)\right \} +h.c. + -i\sum_{u,\nu,i\neq j}J_{u\nu}(\vec{r}_{i}-\vec{r}_{j})[Q,\sigma_{iu}^{\dagger}\sigma_{j\nu}]\\ \nonumber & -i\sqrt{\frac{\gamma_{I}}{\gamma_{0}}}\sum_{u,\nu,i}\left \{ J_{u\nu}(\vec{r}_i-\vec{r}_s) [Q,\sigma_{iu}^{\dagger}s_\nu]+J_{u\nu}(\vec{r}_s-\vec{r}_i)[Q,s_{u}^{\dagger}\sigma_{i\nu}]\right \}+\sum_{u,\nu,i,j}\Gamma _{u\nu}(\vec{r}_i-\vec{r}_j)\left ( \sigma_{iu}^{\dagger}Q\sigma_{j\nu}-\frac{1}{2}{Q,\sigma_{i\nu}^{\dagger}\sigma_{iu}} \right )\\ \nonumber &+\gamma_{I}\left ( s_{u}^{\dagger}Qs_{\nu}-\frac{1}{2}{Q,s_{u}^{\dagger}s_{\nu}} \right )+\sqrt{\frac{\gamma_{I}}{\gamma_{0}}}\left \{ \Gamma_{u\nu}(\vec{r}_i-\vec{r}_s)\left ( \sigma_{iu}^{\dagger}Qs_{\nu}-\frac{1}{2}{Q,\sigma_{iu}^{\dagger}s_{\nu} }\right )+\Gamma_{u\nu}(\vec{r}_s-\vec{r}_i)\left ( s_{u}^{\dagger}Q\sigma_{i\nu}-\frac{1}{2}{Q,s_{u}^{\dagger}\sigma_{i\nu}} \right ) \right \}.
    \end{align}
\end{widetext}
Here,
\begin{align}
    \nonumber J_{u\nu}(\vec{r}_i-\vec{r}_j)&=- \frac{3\pi\gamma_{0}}{\omega_{0}}\hat{d}^{\dagger}_{iu}\cdot \text{Re}[\bar{\bar{G}}(\vec{r}_i-\vec{r}_j,\omega_{0})]\cdot \hat{d}_{jv} \\ \nonumber \Gamma _{u\nu}(\vec{r}_i-\vec{r}_j)&=\frac{6\pi\gamma_{0}}{\omega_{0}}\hat{d}^{\dagger}_{iu}\cdot \text{Im}[\bar{\bar{G}}(\vec{r}_i-\vec{r}_j,\omega_{0})]\cdot \hat{d}_{jv}
\end{align}
Note that $\hat{d}=\vec{d}\setminus |d|$. Additionally, this provides us with specific values for the decay rates $\gamma_{0}=\omega_{0}^3|d|^2\setminus  3\pi\epsilon_{0}$ and $\gamma_{I}=\omega_{0}^3|d_I|^2 \setminus  3\pi\epsilon_{0}$, where $\omega_0 \approx \omega_I$. In the context of an initial state $p_0$, which represents a product state of coherent light with atomic operators and impurity operators, it is possible to separate out the homogeneous field component as follows, 
\begin{align}
    \nonumber
    \vec{E}_{0}^{+}(\vec{r}_i,t)=\left \langle \vec{E}_{0}^{+}(\vec{r}_i,t) \right \rangle+(\vec{E}_{0}^{+}(\vec{r}_i,t)-\left \langle \vec{E}_{0}^{+}(\vec{r}_i,t)\right \rangle).
\end{align} 
Thus we can replace $\vec{E}_{0}^{+}(\vec{r}_i,t)$ with $\left \langle \vec{E}_{0}^{+}(\vec{r}_i,t) \right \rangle$ and define a Langevin noise term F
\begin{widetext}
   \begin{align}
    F=i \sum_{u} \left \{ \sum_{i}\left(\sigma_{iu}^{\dagger}\vec{d}^{\dagger}_{iu}\cdot \left(\vec{E}_{0}^{+}(\vec{r}_i,t)-\left \langle \vec{E}_{0}^{+}(\vec{r}_i,t)\right \rangle\right)\right)+s_{u}^{\dagger}\vec{d}^{\dagger}_{iu}\cdot \left(\vec{E}_{0}^{+}(\vec{r}_s,t)-\left \langle \vec{E}_{0}^{+}(\vec{r}_s,t)\right \rangle\right)\right \}-h.c.,  
    \end{align}
\end{widetext}
such that $\left \langle F\right \rangle=0$. 
Thus, the Langevin noise does not affect any expectation value of the atomic operators $\left \langle Q \right \rangle$, and we omit it for simplicity.
Equations~\ref{eq: appendix_Heisenberg_HLI} and~\ref{eq: appendix_Langevin_QVAF} then results in the total Hamiltonian $H = H_{LI} + H_{AF}$ with an interaction Hamiltonian given by
    \begin{align}
        H_{AF}&=\sum_{u,\nu,i\neq j}J_{u\nu}(\vec{r}_i-\vec{r}_j)\sigma_{iu}^{\dagger}\sigma_{j\nu}\\ \nonumber &+\sqrt{\frac{\gamma_{I}}{\gamma_0}}\sum_{u,\nu,i}\left \{ J_{u\nu}(\vec{r}_i-\vec{r}_s)\sigma_{iu}^{\dagger}s_{\nu}+J_{u\nu}(\vec{r}_s-\vec{r}_i)s_{u}^{\dagger}\sigma_{i\nu} \right \},
    \end{align}
as well as the dissipator    
\begin{widetext}
    \begin{align}
        D[Q]&=\sum_{u,\nu,i,j}\Gamma_{u\nu}(\vec{r}_i-\vec{r}_j)\left ( \sigma_{iu}^{\dagger}Q\sigma_{j\nu}- \frac{1}{2}\{Q,\sigma_{i\nu}^{\dagger}\sigma_{ju}\}\right )+\gamma_{I}\left ( s_{u}^{\dagger}Qs_{\nu}-\frac{1}{2}\{Q,s_{u}^{\dagger}s_{\nu}\}\right )\nonumber \\  &+\sqrt{\frac{\gamma_{I}}{\gamma_{0}}}\sum_{u,\nu,i}\left \{ \Gamma_{u\nu}(\vec{r}_i-\vec{r}_s)\left ( \sigma_{iu}^{\dagger}Qs_{\nu}-\frac{1}{2}\{Q,\sigma_{iu}^{\dagger}s_{\nu}\} \right )+\Gamma_{u\nu}(\vec{r}_s-\vec{r}_i)\left ( s_{u}^{\dagger}Q\sigma_{i\nu}-\frac{1}{2}\{Q,s_{u}^{\dagger}\sigma_{i\nu}\} \right ) \right \}.
    \end{align}
\end{widetext}
This finally results in the equation of motion for the expectation value of operator $Q$
\begin{equation}
    \frac{d}{dt} \langle Q \rangle =  i [H_{LI} + H_{AF},Q]+D[Q]
\end{equation}
Noting that $\langle \hat{O} \rangle = \mathrm{Tr} \{ \hat{O} \hat{\rho}_M \}$, one can readily derive the equation of motion for the atomic density matrix $\rho$, commonly refered to as the master equation
\begin{equation}
    \frac{d}{dt}  \rho  = -i [H_{LI} + H_{AF},\rho]+D^\dagger [\rho],
\end{equation}
where $D^\dagger [\rho]$ indicates the Hermitian conjugate of the dissipator. Notably, $D^\dagger [\rho] = - i [H_{nH},\rho] + D_L[\rho]$ can be split into a non-Hermitian Hamiltonian,
\begin{align}
            H_{nH} &= -i \sum_{u,\nu,i,j} \frac{\Gamma_{u\nu}(\vec{r}_i-\vec{r}_j)}{2} \sigma_{i\nu}^{\dagger}\sigma_{ju} - i \frac{\gamma_{I}}{2} s_{u}^{\dagger}s_{\nu} \nonumber \\ & - i \sqrt{\frac{\gamma_{I}}{\gamma_{0}}}\sum_{u,\nu,i}\left( \frac{\Gamma_{u\nu}(\vec{r}_i-\vec{r}_s)}{2} \sigma_{iu}^{\dagger}s_{\nu}+ \frac{\Gamma_{u\nu}(\vec{r}_s-\vec{r}_i)}{2} s_{u}^{\dagger}\sigma_{i\nu} \right),
\end{align}
that describes the dissipative exchange of an excitation between different atoms or emitters, and a loss-term,
\begin{align}
    D_L[\rho]&=\sum_{u,\nu,i,j} \Gamma_{u\nu}(\vec{r}_i-\vec{r}_j)  \sigma_{iu}\rho\sigma_{j\nu}^{\dagger} +\gamma_{I} s_{u} \rho s_{\nu}^{\dagger}\nonumber \\  &+\sqrt{\frac{\gamma_{I}}{\gamma_{0}}}\sum_{u,\nu,i}\left( \Gamma_{u\nu}(\vec{r}_i-\vec{r}_s) \sigma_{iu} \rho s_{\nu}^{\dagger} +\Gamma_{u\nu}(\vec{r}_s-\vec{r}_i) s_{u} \rho \sigma_{i\nu}^{\dagger} \right),
\end{align}
that describes the loss of an atomic excitation in form of photon emission. In the case where photon emission is heavily suppressed, the action of the loss term can be neglected for times smaller than the inverse effective decay rate. Then, the dynamics of the system can be simply described by a wavefunction and a non-Hermitian Hamiltonian, as discussed in the main text.}
\end{document}